\documentclass[apj]{emulateapj}

\usepackage{bm}
\usepackage{graphicx}
\usepackage{epsf}
\usepackage{graphics}
\usepackage{amsmath}
\usepackage{amssymb}
\usepackage{multirow}
\usepackage{booktabs}
\usepackage{threeparttable}
\usepackage{float}
\usepackage{mathrsfs} 
\DeclareMathSizes{15}{1}{1}{1}

\newcommand{\msunh}{\>h^{-1}\rm M_\odot}

\newcommand{\mpch}{\>h^{-1}{\rm {Mpc}}}

\newcommand{\kmsmpc}{\>{\rm km}\,{\rm s}^{-1}\,{\rm Mpc}^{-1}}

\newcommand{\rmag}{\>^{0.1}{\rm M}_r-5\log h}

\newcommand{\rmd}{{\rm d}}
\newcommand{\rmnp}{{\rm p}}
\newcommand{\rmm}{{\rm m}}
\newcommand{\rmh}{{\rm h}}

\newcommand{\xis}{{\xi(s)}}
\newcommand{\xipi}{{\xi(r_{\rm p},r_\pi)}}
\newcommand{\xipimod}{{\xi_{\rm mod}(r_{\rm p},r_\pi)}}
\newcommand{\xipimea}{{\xi_{\rm mea}(r_{\rm p},r_\pi)}}
\newcommand{\ximod}{{\xi_{\rm mod}}}
\newcommand{\ximea}{{\xi_{\rm mea}}}
\newcommand{\fo}{{f(\Omega_\rmm)}}

\newcommand{\fss}{f\sigma_8 =0.376 \pm 0.038}

\newcommand{\f}{$0.464^{+0.040}_{-0.040}$}
\newcommand{\s}{$0.769^{+0.121}_{-0.089}~$}
\newcommand{\bI}{$1.910^{+0.234}_{-0.268}$}
\newcommand{\bII}{$1.449^{+0.194}_{-0.196}$}
\newcommand{\bIII}{$1.301^{+0.170}_{-0.177}$}
\newcommand{\bIV}{$1.196^{+0.159}_{-0.161}~$}  

\def\gtsima{$\; \buildrel > \over \sim \;$}
\def\ltsima{$\; \buildrel < \over \sim \;$}
\def\prosima{$\; \buildrel \propto \over \sim \;$}
\def\gsim{\lower.7ex\hbox{\gtsima}}
\def\lsim{\lower.7ex\hbox{\ltsima}}
\def\simgt{\lower.7ex\hbox{\gtsima}}
\def\simlt{\lower.7ex\hbox{\ltsima}}
\def\simpr{\lower.7ex\hbox{\prosima}}
\def\la{\lsim}

\def\lta{\la}

\usepackage{color}

\shorttitle{Mapping the real space distributions of galaxies}
\shortauthors{Shi et al.}

\begin{document}

\title{Mapping the Real Space Distributions of Galaxies in SDSS DR7:
  II.  Measuring the growth rate, clustering amplitude of matter 
  and biases of galaxies at redshift $0.1$}

\author{Feng Shi\altaffilmark{1,8}, Xiaohu Yang\altaffilmark{2,3},
  Huiyuan Wang\altaffilmark{4}, Youcai Zhang\altaffilmark{1},
  H.J. Mo\altaffilmark{5,6}, Frank C. van den Bosch\altaffilmark{7},
  Wentao Luo\altaffilmark{2} Dylan Tweed\altaffilmark{2}, Shijie
  Li\altaffilmark{2}, Chengze Liu\altaffilmark{2}, Yi
  Lu\altaffilmark{1}, Lei Yang\altaffilmark{2} }

\altaffiltext{1}{Shanghai Astronomical Observatory, Nandan Road 80,
  Shanghai 200030, China; E-mail: sfeng@shao.ac.cn}

\altaffiltext{2}{Department of Astronomy, Shanghai Key Laboratory for
  Particle Physics and Cosmology, Shanghai Jiao Tong University,
  Shanghai 200240, China; E-mail: xyang@sjtu.edu.cn}

\altaffiltext{3}{IFSA Collaborative Innovation Center, and Tsung-Dao
  Lee Institute, Shanghai Jiao Tong University, Shanghai 200240,
  China}

\altaffiltext{4}{Key Laboratory for Research in Galaxies and Cosmology,
University of Science and Technology of China, Hefei, Anhui 230026, China}

\altaffiltext{5}{Physics Department and Center for Astrophysics,
  Tsinghua University, Beijing 10084, China}

\altaffiltext{6}{Department of Astronomy, University of Massachusetts,
Amherst MA 01003-9305, USA}

\altaffiltext{7}{Department of Astronomy, Yale University, P.O. Box 208101,
  New Haven, CT 06520-8101, USA}

\altaffiltext{8}{University of Chinese Academy of Sciences, 19A,
  Yuquan Road, Beijing, China}

\begin{abstract}
  We extend the real-space mapping method developed in \citet{Shi16}
  so that it can be applied to flux-limited galaxy samples.  We use an
  ensemble of mock catalogs to demonstrate the reliability of this
  extension, showing that it allows for an accurate recovery of the
  real-space correlation functions and galaxy biases. We also
  demonstrate that, using an iterative method applied to intermediate-scale 
  clustering data, we can obtain an unbiased estimate of the
  growth rate of structure $f\sigma_8$, which is related to the
  clustering amplitude of matter, to an accuracy of $\sim 10\%$.
  Applying this method to the Sloan Digital Sky Survey (SDSS) Data
  Release 7 (DR7), we construct a real-space galaxy catalog spanning
  the redshift range $0.01 \leq z \leq 0.2$, which contains 584,473
  galaxies in the north Galactic cap (NGC). Using this data, we infer
  $\fss$ at a median redshift $z=0.1$, which is consistent with the
  WMAP9 cosmology at the $1\sigma$ level. By combining this measurement
  with the real-space clustering of galaxies and with galaxy-galaxy
  weak lensing measurements for the same sets of galaxies, we are able
  to break the degeneracy between $f$, $\sigma_8$, and $b$. From the
  SDSS DR7 data alone, we obtain the following cosmological
  constraints at redshift $z=0.1$: $f=$\f , $\sigma_8=$\s, and $b=$
  \bI, \bII, \bIII, and \bIV for galaxies within different absolute
  magnitude bins
  $\rmag=[-23,0, -22.0], [-22,0, -21.0], [-21.0, -20.0]$ and
  $[-20.0, -19.0]$, respectively.
\end{abstract}

\keywords {cosmology: observation - cosmology: large-scale structure of universe -
 galaxies: distances and redshifts - methods: statistical}

\section{Introduction}
\label{sec:intro}

High-precision measurements of the growth of structure are required to
understand the nature of the accelerating expansion of the universe,
which can be explained by either dark energy or modified gravity
\citep[e.g.][]{Amen2005, Jain2008, Linder2008, Wang2008, Per2009,
  Song2009, White2009, Jen2011, Cai2012}.  One of the most powerful
tools to perform this measurement is redshift-space distortions 
\citep[RSD; e.g.][]{Sar1977, Dav1983, Kai1987, Reg1991, Ham1992, Wey1993},
which give rise to an anisotropic two-point correlation function (2PCF)
in redshift space.

These anisotropies arise because redshifts include both the Hubble
expansion and the peculiar velocity of the galaxies along the
line of sight.  The magnitude of the anisotropies therefore depends on
the amplitude of the velocity field, which is commonly parameterized,
on large scales, by $f \sigma_8$, where $f = \rmd\ln D/\rmd\ln a$ is
the logarithmic derivative of the linear growth factor, $D$, with
respect to the scale factor, $a$, and $\sigma_8$ is the
clustering amplitude of matter.  In general, we have that
$f = \Omega_\rmm(z)^{\gamma}$, with $\Omega_\rmm(z)$ the matter
density parameter at redshift $z$, and, in the case of General
Relativity (GR), $\gamma \simeq 0.55$ \citep[e.g.][]{LinderCahn2007}.
Hence, we can use the redshift evolution of $f\sigma_8$ to test our
law of gravity \citep{Song2009}. In addition, since the redshift
evolution of the linear growth rate depends on the equation of state
of dark energy, $f\sigma_8$ can also be used to constrain the nature
of dark energy. This method has been applied successfully to data from
6dFGS \citep{Beu2012}, WiggleZ \citep{Blake2011}, VIPERS
\citep{Torre2013}, and the SDSS \citep[e.g.][]{Chuang2013, Beu2014,
  Oka2014, Reid2014, Sam2014, Alam2015, How2015}.

The overall clustering amplitude of galaxies depends on both
$\sigma_8$ and the galaxy bias parameter $b$, to the extent that an
observed galaxy correlation function constrains the product
$b\sigma_8$. By taking a ratio of the quadrupole and monopole terms of
the 2PCF in redshift space, one obtains a
measure for the RSD, $\beta = f/b$, that is independent of the
power-spectrum normalization, $\sigma_8$. Hence, if one could
independently constrain the galaxy bias, one could use this ratio, and
thus the RSD, to constrain the linear growth rate.  Alternatively, one
can measure $f\sigma_8$ \citep{Song2009} without facing the difficulty
of measuring the galaxy bias. Note that since the nonlinear redshift
distortion effect, also know as the finger-of-God (FOG) effect, can impact the
clustering pattern to quite large scales, in order to have an unbiased
constraint on $f\sigma_8$, one needs to use the clustering measurements
on very large scales. However, since the 2PCFs on large scales are close 
to zero and noisy, it is not easy to obtain reliable and accurate 
constraints on $f\sigma_8$ 
\citep[but see][for such a probe in Fourier space]{Li2016}.

In this paper, we present a method that can simultaneously measure the
real-space 2PCF $\xi(s)$ (including $b\sigma_8$) and constrain
$f\sigma_8$ using intermediate-scale clustering measurements.  In
\citet[][hereafter S16]{Shi16}, the first paper in this series, we
developed a method to correct RSD for individual galaxies, and
used it to construct the real-space distribution of galaxies in the
SDSS DR7. S16 mainly presented measurements of the real-space
2PCF, and the bias relative to the
underlying matter distribution for galaxies of different luminosities
and colors.  Here we improve upon the reconstruction method of S16 by
using all data from the flux-limited SDSS galaxy sample, rather
than restricting the method to volume-limited
subsamples. We use this new and improved method to measure the growth
rate parameter $f\sigma_8$. Since the reconstruction is
cosmology-dependent, assuming an incorrect cosmology results in
systematic errors in our velocity reconstruction and thus in
distortions (i.e., residual anisotropies) in the correlation
functions. We can therefore use the performance of redshift distortion
correction to constrain cosmological parameters. The main advantage of
this method is that it can provide a measure of the real-space 2PCF
$\xi(s)$ (including $b\sigma_8$) and linear growth rate $f\sigma_8$ in
an unbiased way. Moreover, when combined with galaxy-galaxy lensing
shear measurements, one can disentangle the degeneracies among these
parameters and provide individual constraints on $f$, $\sigma_8$, and
$b$.

This paper is organized as follows. In Section~\ref{sec:data} we
present the galaxy and group catalogs used in this paper and
introduce the methods to correct for the RSDs.
We use mock samples to test the reliability of our correction model in
Section~\ref{sec:validation}. In Section~\ref {sec:estimation} we
describe our method for constraining the growth rate of structure and
test its reliability using mock samples.  In Section
\ref{sec:application} we apply our method to the SDSS DR7 to constrain
$f$, $\sigma_8$ and $b$. Finally, we summarize our main findings in
Section~\ref {sec:con}. Throughout this paper, unless stated
otherwise, we adopt a fiducial $\Lambda$CDM cosmological model with
WMAP9 parameters \citep{Hin2013}: $\Omega_\rmm = 0.282$,
$\Omega_{\Lambda} = 0.718$, $\Omega_{\rm b} = 0.046$, $n_{\rm
  s}=0.965$, $h=H_0/(100 \kmsmpc) = 0.697$ and $\sigma_8 = 0.817$.

\section{OBSERVATIONAL DATA}
\label{sec:data}

\subsection{Galaxy group catalog}

Our sample of galaxies is taken from the New York University
Valued-Added Galaxy Catalog \citep[NYU-VAGC;][]{Bla2005}. This catalog
is based on the SDSS DR7 \citep{Aba2009}, with an independent set of
significantly improved reduction algorithms over the original
pipeline.  Our analysis is based on galaxies in the main galaxy sample
with extinction-corrected apparent magnitudes brighter than $r=17.72$,
within the redshift range $0.01 \leq z \leq 0.20$, and with redshift
completeness $ C_z > 0.7$. In S16, we used these data to reconstruct the
velocity field and correct for RSDs, using a volume-limited sample
that contains only 396,068 galaxies in the northern Galactic cap (NGC)
with redshifts in the range $0.01 \leq z \leq 0.12$. In order to make
full use of the data available, here we extend the method to a
flux-limited sample and apply the reconstruction to all galaxies in
the contiguous NGC region, which consists of 584,473 galaxies covering
$7047$ deg$^2$ on the sky.  The median redshift of the sample is at
$z_{\rm med} = 0.1$. Finally, using this sample, we construct
flux-limited subsamples for galaxies in the following four absolute
$r$-band magnitude bins: $\rmag=$ $\bm{[}-23.0,-22.0\bm{]}$,
$\bm{[}-22.0,-21.0\bm{]}$, $\bm{[}-21.0,-20.0\bm{]}$, and
$\bm{[}-20.0,-19.0\bm{]}$. The corresponding redshift ranges, numbers
of galaxies, and average magnitudes are listed in Table~\ref{subsamp}.

A key ingredient of our method for reconstructing the real-space
distribution of galaxies (see \S\ref{sec:correction}) is	 galaxy
groups.  As in S16, we make use of the SDSS DR7 group catalog of
\citet{Yang2012}, constructed using the adaptive halo-based group
finder developed by \cite{Yang2005, Yang2007} and updated to the WMAP9
cosmology adopted here. The \citeauthor{Yang2005} group finder is
optimized to group galaxies that reside in the same dark matter host
halo. Halo masses are assigned to each group using the ranking of
either their total characteristic luminosity or the total
characteristic stellar mass. These are computed using all group
member galaxies more luminous than $^{0.1}M_r - 5 \log h = -19.5$. As
demonstrated in \cite{Yang2007}, these two estimates of halo mass
agree very well with each other. Here, as in S16, we adopt the halo
masses based on the characteristic luminosity ranking.
\begin{table}
\center
\scalebox{0.85}{
\begin{threeparttable}[c]
\caption{Galaxy Flux-limited Subsamples }\label{subsamp}
\setlength{\tabcolsep}{3pt}
\begin{tabular}{ccccc}
\toprule

\multirow{2}{*}{Sample ID}  &
\multirow{2}{*}{$\rmag$ }  &
\multirow{2}{*}{Redshift}  & 
\multirow{2}{*}{~~~$N_{\rm gal}$}  &
\multirow{2}{*}{Averaged Magnitude} \\ \\
(1) & (2) & (3) & (4) & (5) \\

\hline
\multirow{2}{*}{$1$} &
\multirow{2}{*}{$\bm{[}-23,-22\bm{]}$} & 
\multirow{2}{*}{$\bm{[}0.01, 0.20\bm{]}$}  & 
\multirow{2}{*}{$~~~10,340$} &
\multirow{2}{*}{$-22.22$} \\

\multirow{2}{*}{$2$} &
\multirow{2}{*}{$\bm{[}-22,-21\bm{]}$} & 
\multirow{2}{*}{$\bm{[}0.01, 0.20\bm{]}$}  & 
\multirow{2}{*}{$~~~150,030$} &
\multirow{2}{*}{$-21.34$} \\

\multirow{2}{*}{$3$} &
\multirow{2}{*}{$\bm{[}-21,-20\bm{]}$} & 
\multirow{2}{*}{$\bm{[}0.01, 0.20\bm{]}$}  & 
\multirow{2}{*}{$~~~229,104$}  &
\multirow{2}{*}{$-20.44$}  \\

\multirow{2}{*}{$4$} &                                       
\multirow{2}{*}{$\bm{[}-20,-19\bm{]}$} & 
\multirow{2}{*}{$\bm{[}0.01, 0.20\bm{]}$}  & 
\multirow{2}{*}{$~~~127,490$} &
\multirow{2}{*}{$-19.57$}  \\
 
\\ \bottomrule
\end{tabular}
\textbf{Notes.} Columns (1)-(5) correspond to the ID number,
  absolute magnitude range, redshift range, number of galaxies and the
  averaged absolute magnitude for each galaxy sample, respectively.
\end{threeparttable}}
\end{table}

\subsection{Correcting for RSDs}
\label{sec:correction}

In the survey, since galaxy redshifts are not exact measures of 
distances, the observed galaxy distribution is distorted with respect
to the true distribution. The observed redshift $z_{\rm obs}$, related
to the redshift distance, consists of a cosmological redshift 
$z_{\rm cos}$, arising from the Hubble expansion plus a Doppler 
contribution due to the line-of-sight component of the galaxy's peculiar 
velocity $v_{\rm pec}$. Peculiar velocities thus lead to RSDs, 
which contain important information regarding the 
growth of structure in our universe. The RSDs have different observational
consequences on different scales, such as the small-scale FOG
effect \citep{Jac1972, Tul1978} and the large-scale Kaiser effect 
\citep{Kai1987}. Generally, the FOG effect is caused by the 
nonlinear virialized motions of galaxies within dark matter halos, while
the Kaiser effect is by the linear infall motions of galaxies toward 
overdense regions.

Actually, the peculiar velocity of a galaxy can be split into
two components:
\begin{equation}
v_{\rm pec} = v_{\rm cen} + v_{\sigma}\,.
\end{equation} 
Here $v_{\rm cen}$ is the center velocity of the halo in which
the galaxy resides, and $v_{\sigma}$ is the velocity of the galaxy 
with respect to that halo center. Note that the velocities are both
along line of sight. Roughly speaking, $v_{\rm cen}$ 
contributes to the Kaiser effect, while $v_{\sigma}$ contributes mainly 
to the FOG effect. In our method, it is then useful to correct for the 
Kaiser and FOG effects separately.

In order to correct for the Kaiser effect, we reconstruct the velocity
field in the linear regime using the method of \citet[][hereafter
W12]{WH2012}.  Here we briefly summarize the main ingredients of this
reconstruction method and refer the reader to W12 for more details.
In the linear regime, the peculiar velocities are induced by and
proportional to the perturbations in the matter distribution. In
Fourier space, we have 
\begin{equation}\label{eq_vk}
\bm{v}(\bm{k}) = H \, a \, f(\Omega) \, \frac{i\bm{k}}{k^2} \, \delta(\bm{k}).
\end{equation}
Here $H = \dot{a}/a$ is the Hubble parameter, $a$ is the scale factor,
and $\delta(\bm{k})$ is the Fourier transform of the density
perturbation field $\delta(\bm{x})$. Hence, for a given cosmology, one
can directly infer the linear velocity field from the density
perturbation field, $\delta(\bm{x})$. Meanwhile, as
$\delta(\bm{x})=\sigma_8 \delta_{\sigma_8=1}(\bm{x})$, we can write
\begin{equation}\label{eq_vk2}
  \bm{v}(\bm{k}) = H \, a \, f(\Omega) \, \sigma_8 \, \frac{i\bm{k}}{k^2} \, \delta_{\sigma_8=1}(\bm{k}),
\end{equation}
which indicates that at a given redshift, the amplitude of the velocity
field is, to first order, linearly proportional to $f(\Omega) \sigma_8$.

In practice, the peculiar velocity field is reconstructed from 
the halo density field by replacing $\delta(\bm{k})$ in Eq.~(\ref{eq_vk}) 
with $\delta_\rmh(\bm{k}) / b_{\rm hm}$, where $\delta_\rmh$ is the 
dark matter halo density field and $b_{\rm hm}$ is the linear bias 
parameter for dark matter halos with mass $M_\rmh \geq M_{\rm th}$, 
which is given by
\begin{equation}\label{bhm}
  b_{\rm hm} = {\int_{M_{\rm th}}^{\infty} M \, b_\rmh(M) \, n(M) \, {\rm d}M \over
\int_{M_{\rm th}}^{\infty} M \, n(M) \, {\rm d}M}.  
\end{equation}
Here $n(M)$ and $b_\rmh(M)$ are the halo mass function and the halo
bias function, respectively. In other words, the velocity field
can be reconstructed even from a limited distribution of dark
matter halos above some mass threshold. We can then actually extract the 
latter from our galaxy group catalog in a fairly straightforward manner.

In S16 we used a volume-limited galaxy group sample with $M_\rmh \geq
M_{\rm th} = 12.5\msunh$ and redshift $z \leq 0.12$.  However, the
volume-limited sample excludes many galaxies and greatly limits the
sensitivity to large-scale modes. In this paper, we improved upon this
by using a flux-limited galaxy group sample instead.  This adds one
nontrivial complication, though: in flux-limited samples, $b_{\rm
  hm}$ is no longer a constant like in a volume-limited sample but rather
a function of the redshift $z$. In order to take this into account, we
divide the SDSS volume into six subvolumes (or redshift bins). Each
subvolume has its own mass threshold, $M_{\rm th}$, which we use to
compute the corresponding bias parameter, $b_{\rm hm}$, using
Eq.~(\ref{bhm}) and adopting the halo mass and halo bias functions of
\citet{Tinker2008}. The mass threshold, $M_{\rm th}$, is obtained from
the halo mass below which the halo mass distribution starts to drop
systematically. Table~\ref{Subvol} lists the redshift range, mass
threshold $M_{\rm th}$ and bias $b_{\rm hm}$ for our six subvolumes.

Next, we embed the six subvolumes in a periodic cubic box of 1111
$\mpch$ on a side, divide the box into $1024^3$ grid cells, and
compute $\delta(\bm{x}) = \delta_\rmh(\bm{x})/b_{\rm hm}$ on that
grid, where the value of $b_{\rm hm}$, listed in Table~\ref{Subvol},
is selected depending on which subvolume the halo is located in.
Note that the location of each group is defined as the luminosity-weighted 
center of all group members.  Next, we smooth
$\delta(\bm{x})$ using a Gaussian smoothing kernel with a mass scale
of $10^{14.75}\msunh$, and fast Fourier transform this smoothed
over-density field to compute $\bm{v}(\bm{k})$ using
Eq.~(\ref{eq_vk}). The velocity field of the group centers $\bm{v}_{\rm
  cen}(\bm{x})$ is simply estimated from the Fourier transform of
$\bm{v}(\bm{k})$.  Finally, we compute, for each galaxy, the
Kaiser-corrected redshift as
\begin{equation}\label{zcorr}
z_{\rm corr} = {z_{\rm obs} - (v_{\rm cen}/c) \over 1 + (v_{\rm cen}/c)}\,.
\end{equation}
Since the velocity field is computed using the redshift-space
distribution of the groups, this method needs to be iterated until
convergence is achieved. As \citet{WH2009, WH2012} suggested, 
two iterations are generally sufficient.

Next, we move to the correction for the FOG effect. As discussed  
previously, we focus on correcting for RSDs using the flux-limited sample
in this paper instead of the volume-limited sample in S16. Actually,
the main difference of correction between the two samples is in 
reconstructing the velocity field to correct for the Kaiser effect, while 
in the FOG correction, the method is fully the same for the two kinds of samples.
Here we briefly summarize the main ingredients of the method 
and refer the reader to S16 for more details. 

We correct for the FOG effect in a statistical sense, with the 
assumption that group galaxies are unbiased tracers of the halo's mass 
distribution and therefore follow an NFW \citep{NFW1997} radial 
number density profile. In practice, we do not displace central 
galaxies and just assign the satellites new positions in the group
by randomly drawing a line-of-sight distance, $r_{\pi}$, 
for satellites whose probability follows the NFW profile with 
$r=\sqrt{r_\rmnp^2+r_{\pi}^2}$. Here $r_\rmnp$ is the projected 
distance between the satellite and the luminosity-weighted center 
of its group. Although the FOG correction is model-dependent, it is 
useful to recover the large-scale clustering of galaxies. As we will 
discuss below, the FOG effect caused by the small-scale velocities 
also has a significant effect on the large-scale clustering of galaxies. 
Meanwhile, it is also a necessary step to constrain the growth of 
structure in our method.
 
Finally, the galaxy is assigned a comoving distance 
given by $r(z_{\rm corr})+ r_{\pi}$, where $z_{\rm corr}$ is given 
by Eq.~(\ref{zcorr}). 
Our method therefore consists of the following four steps.
\begin{enumerate}
\item Assigning a halo mass to each group based on its characteristic
  luminosity, where the groups are constructed using a halo-based
  group finder in redshift space.
\item Correcting, in a statistical sense, for the FOG effect by
  randomly assigning new line-of-sight positions to satellite
  galaxies.  It is assumed that satellite galaxies follow an NFW radial
  number density distribution within their host halos.
\item Correcting for the Kaiser effect using the velocity field
  reconstructed from the biased halo density field with bias estimated
  in the flux-limited sample.
\item Computing for each galaxy the corrected redshift and
  corresponding comoving distance.
\end{enumerate}
Although this is the order in which we apply our method, we point out
that it makes no difference whether one first applies the FOG
correction followed by the Kaiser correction, or vice versa.

Finally, S16 has defined a number of different spaces according
to what kind of velocity ($v_{\rm cen}$, $v_{\sigma}$ , $v_{\rm pec}$) 
is used in computing the redshift of the galaxy. Here we also give a brief 
description of the various spaces in Table~\ref{spaces} for completeness. 
In what follows, the top four spaces are referred to as `true' spaces, 
which are based on true velocities and true groups (dark matter halos)
without observational errors or errors in group identifications and/or 
membership. The bottom three spaces are reconstructed spaces, obtained 
by correcting for the corresponding redshift distortions, such as 
re-Kaiser space in which only the FOG effect is corrected, the re-FOG space 
in which only the Kaiser effect is corrected, and the re-real space in which 
both corrections are applied. These are based on the reconstructed velocity 
field and on groups identified by applying the group finder in redshift space.

\begin{table}
\center
\scalebox{1.0}{
\begin{threeparttable}[c]
\caption{Subvolumes in Reconstruction }\label{Subvol}
\setlength{\tabcolsep}{14pt}
\begin{tabular}{cccccc}
\toprule
\multirow{2}{*}{Redshift Range} &
\multirow{2}{*}{Mass Threshold} & 
\multirow{2}{*}{Bias} \\

\multirow{2}{*}{$z$} &
\multirow{2}{*}{$M_{\rm th}$} & 
\multirow{2}{*}{$b_{\rm hm}$} \\\\

(1) & (2) & (3) \\

\hline
\multirow{2}{*}{$\bm{[}0.010, 0.083\bm{]}$}  & 
\multirow{2}{*}{$11.67$} &
\multirow{2}{*}{$1.50$} \\

\multirow{2}{*}{$\bm{[}0.083, 0.115\bm{]}$}  & 
\multirow{2}{*}{$11.98$} &
\multirow{2}{*}{$1.57$} \\

\multirow{2}{*}{$\bm{[}0.115, 0.135\bm{]}$}  & 
\multirow{2}{*}{$12.28$} &
\multirow{2}{*}{$1.66$} \\

\multirow{2}{*}{$\bm{[}0.135, 0.150\bm{]}$}  & 
\multirow{2}{*}{$12.59$} &
\multirow{2}{*}{$1.77$} \\

\multirow{2}{*}{$\bm{[}0.150, 0.179\bm{]}$}  & 
\multirow{2}{*}{$12.89$} &
\multirow{2}{*}{$1.92$} \\

\multirow{2}{*}{$\bm{[}0.179, 0.200\bm{]}$}  & 
\multirow{2}{*}{$13.20$} &
\multirow{2}{*}{$2.12$} \\

\\ \bottomrule
\end{tabular}
\textbf{Notes.} Column (1) lists the redshift range for each of
  the six subvolumes. Listed in column (2), $M_{\rm th}$, is the halo
  mass threshold to which the sample is complete in each
  subvolume. In column (3), $b_{\rm hm}$ is the linear bias parameter
  for a halo with mass $M_\rmh \geq M_{\rm th}$, which is computed
  according to Eq.(\ref{bhm}).
\end{threeparttable}}
\end{table}

\begin{table*}
\scalebox{1.2}{
\begin{threeparttable}
\caption{Description of different spaces.}\label{spaces}
\centering

\setlength{\tabcolsep}{20pt}
\begin{tabular}{ll}
\toprule
Space     &  Description \\
\hline
\multirow{2}{*}{Real space} & 
\multirow{2}{*}{Survey geometry without redshift distortions}  \\
\\ \hline
\multirow{2}{*}{FOG space} & 
\multirow{2}{*}{Distorted only by FOG effect:
 ~~$z_{\rm obs}=z_{\rm cos}+\frac{v_{\sigma}}{c}(1+z_{\rm cos})$ } \\
\\ \hline
\multirow{2}{*}{Kaiser space} & 
\multirow{2}{*}{Distorted only by Kaiser effect:
 ~~$z_{\rm obs}=z_{\rm cos}+\frac{v_{\rm cen}}{c}(1+z_{\rm cos})$} \\
\\ \hline
\multirow{2}{*}{Redshift space} & 
\multirow{2}{*}{Distorted by both Kaiser and FOG effects:
 ~~$z_{\rm obs}=z_{\rm cos}+\frac{v_{\rm pec}}{c}(1+z_{\rm cos})$ } \\
\\ \toprule
\multirow{2}{*}{Re-real space} & 
\multirow{2}{*}{Reconstructed real space; based on correcting RSDs } \\
\\ \hline
\multirow{2}{*}{Re-Kaiser space} & 
\multirow{2}{*}{Reconstructed Kaiser space; based on correcting for FOG effect only } \\
\\ \hline
\multirow{2}{*}{Re-FOG space} & 
\multirow{2}{*}{Reconstructed FOG space; based on correcting for Kaiser effect only } \\
\\ \bottomrule
\end{tabular}
\textbf{Notes.} The first four spaces are `true' spaces based on true
groups (all galaxies belonging to the same dark matter halo). The
final three spaces are `reconstructed' spaces based on groups
identified by applying the group finder in redshift space.
\end{threeparttable}}
\end{table*}

\begin{figure*}
\center
\includegraphics[width=0.8\textwidth]{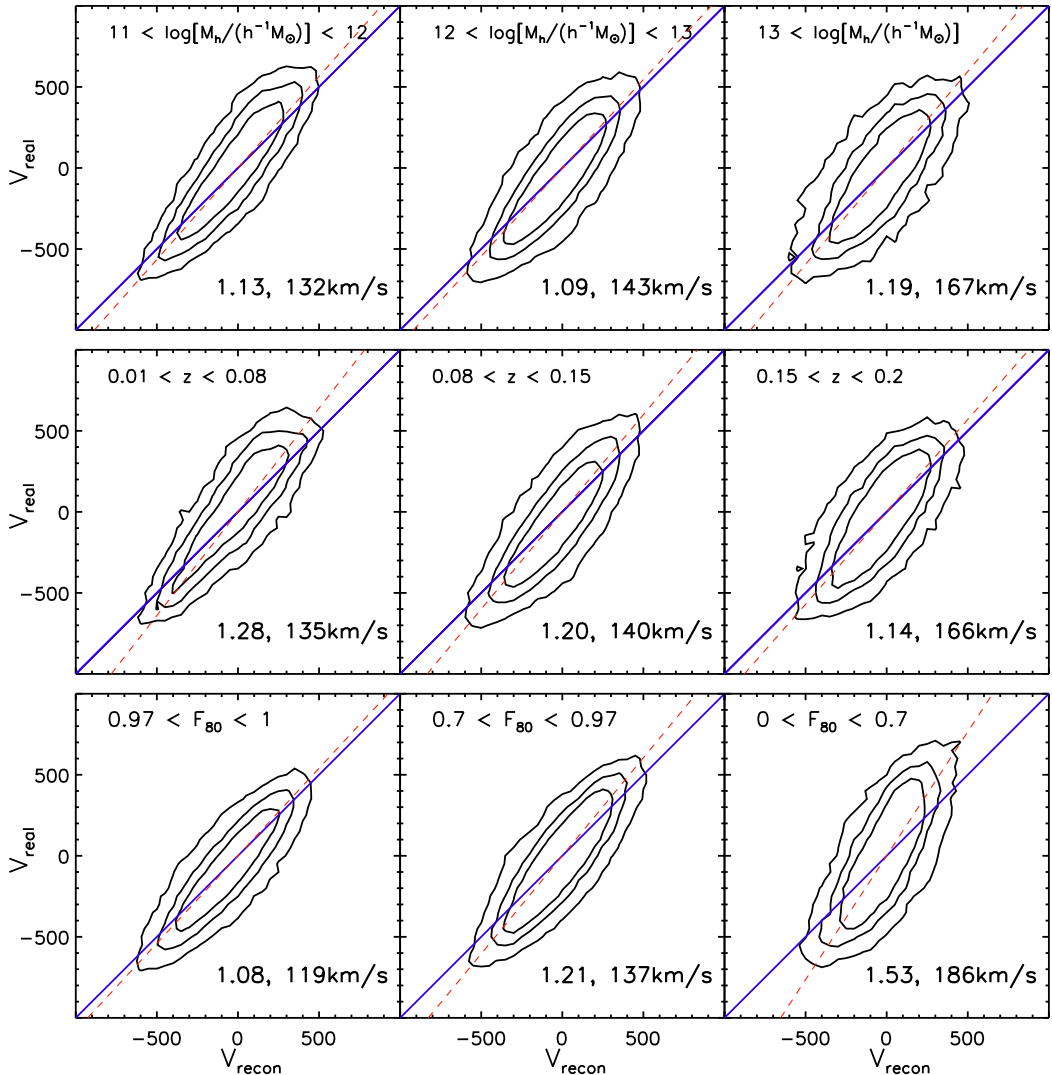}
\caption{Validations with mocks: true group velocities, $v_{\rm
    real}$, obtained directly from the mock vs. the corresponding
  reconstructed velocity, $v_{\rm recon}$, obtained by applying our
  reconstruction method to the mock data in redshift space.  Note that
  all of these velocities are along the line of sight.  The top and
  middle rows show results for galaxies in different halo mass and
  redshift bins, respectively, while the panels in the bottom row show
  results for different bins in the filling factor, $F_{80}$. The
  contours in each panel encompass $50\%$, $70\%$, and $85\%$ per cent of
  the groups in each subsample. The slope of the best-fitting relation
  and the scatter, in terms of the rms in $v_{\rm real}-v_{\rm recon}$,
  are also indicated in each panel.}
\label{fig:vel}
\end{figure*}  
\begin{figure*}
\center
\includegraphics[width=0.9\textwidth]{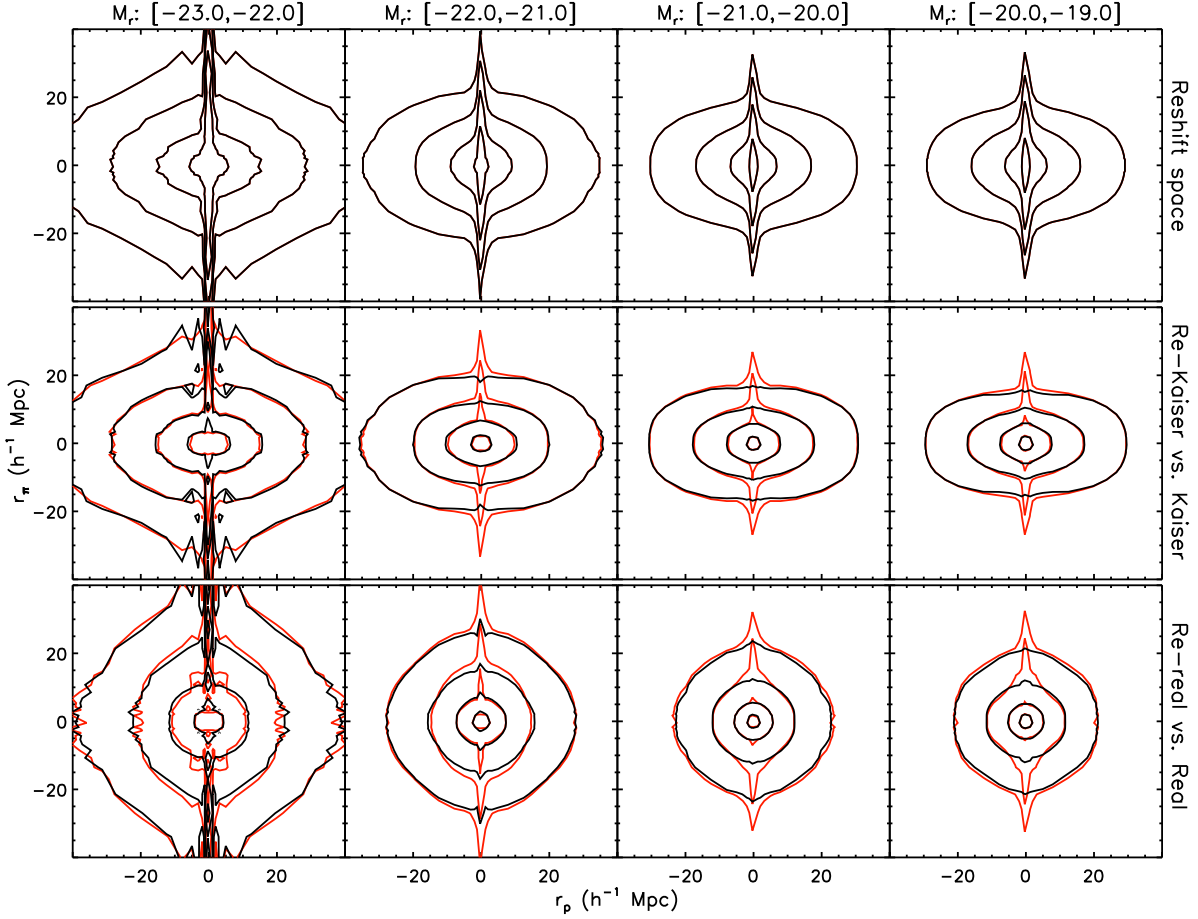}
\caption{Validations with mocks: comparison of two-dimensional 2PCFs
  of mock galaxies. Different columns correspond to mock galaxies in
  different absolute $r$-band magnitude bins, as indicated at the top
  of each column. Different rows correspond to different spaces, as
  indicated at the right of each row. Black and red contours
  correspond to the results in the true and reconstructed spaces,
  respectively, with contour levels corresponding to
  $\xi=5,~1,~0.3,~0.1$. }
\label{fig:2d2pcf}
\end{figure*}    
\begin{figure*}
\center
\includegraphics[width=0.85\textwidth]{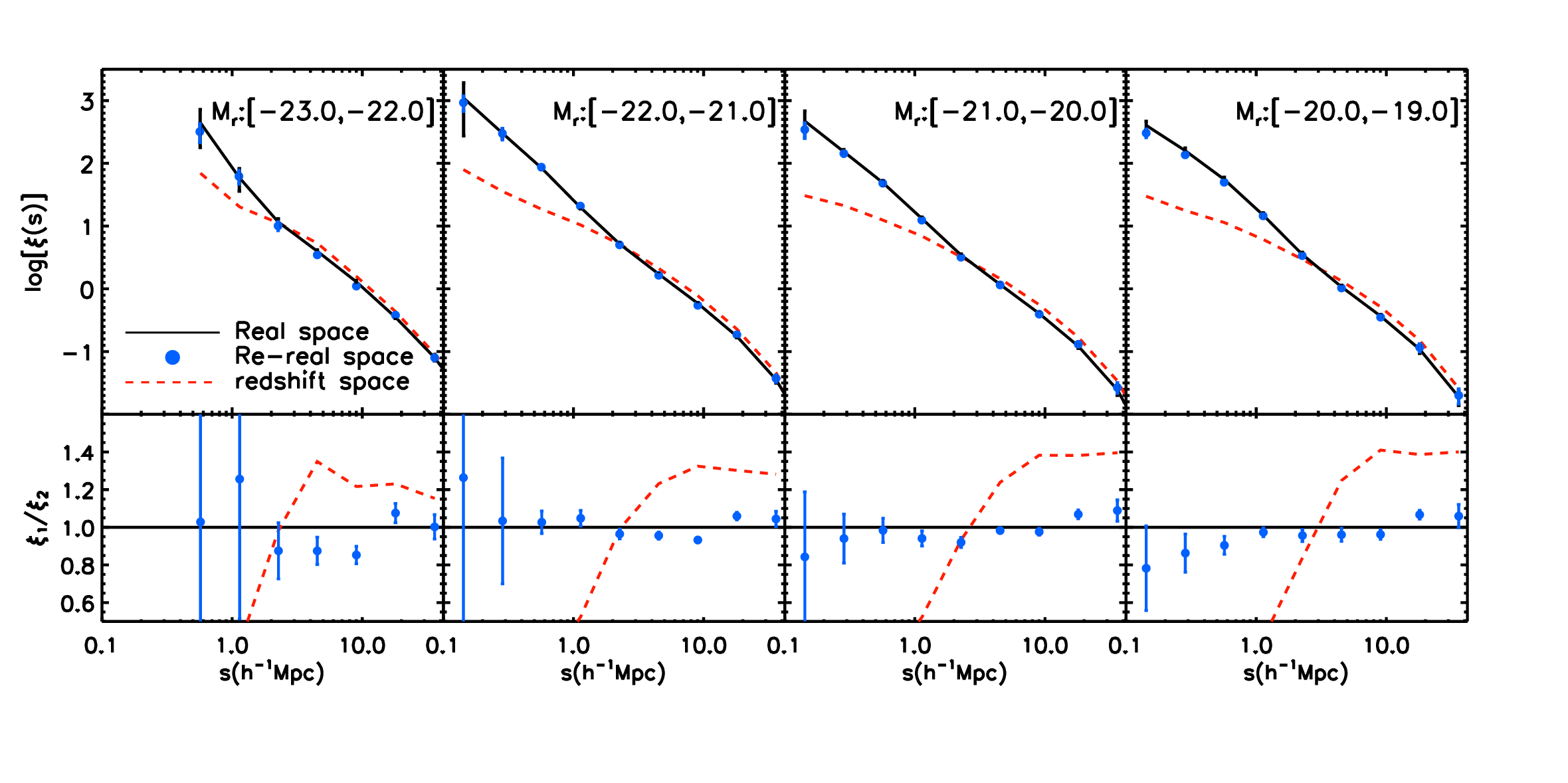}
\caption{Validations with mocks: 2PCFs (upper panels) and 2PCF ratios
  (lower panels) for mock galaxies in real vs. re-real space. The
  solid line in the upper panels indicates the 2PCF in the real space,
  averaged over 10 mock samples, while the blue filled circles indicate
  the corresponding average 2PCF in the re-real space, with the error
  bars indicating the $\pm 1 \sigma$ variance among the 10 mock
  samples. The red dashed lines indicate the corresponding 2PCFs in
  redshift space and are shown for comparison.  The lower panels plot
  the average and $\pm 1 \sigma$ variance of the ratio of the 2PCFs in
  the re-real space over that in the real space (blue filled circles
  with error bars). The red dashed lines indicate the ratio of the
  redshift-space 2PCF to the real-space one. Different columns
  correspond to different $r$-band magnitude bins, as
  indicated.}\label{fig:xiS}
\end{figure*}  

\section{Validation with Mock Data}
\label{sec:validation}

In order to test and validate the method described above, we first
apply it to a mock SDSS DR7 galaxy catalog. Although W12 and S16
already presented several tests regarding the reconstruction of the
velocity field and the real-space correlation function, here we focus
specifically on testing the application of our reconstruction method
to a flux-limited sample.

\subsection{Mock Catalogs}
\label{sec:mocks}

The mocks that we use here are exactly the same as those used in S16.
For completeness, though, we briefly describe the main ingredients in
what follows. The mocks are constructed from a high-resolution N-body
simulation that evolves the distribution of $3072^3$ dark matter
particles in a periodic box of $500 \mpch$ on a side
\citep{Li2016}. This simulation was carried out at the Center for High
Performance Computing at Shanghai Jiao Tong University and was run
with {\tt L-GADGET}, a memory-optimized version of {\tt GADGET2}
\citep{Spr2005}. The cosmological parameters adopted by this
simulation are consistent with the WMAP9 results \citep{Hin2013}.
Dark matter halos are identified using the standard friends-of-friends
(FoF) algorithm \citep[e.g.][]{Dav1985} with a linking length that is
0.2 times the mean interparticle separation. Mock galaxies are
assigned to dark matter halos using the conditional luminosity
function \citep[hereafter CLF, see][]{Yang2003} as constrained by
\citet{Cac2013}. The algorithm used to assign luminosities and
phase-space coordinates to the mock galaxies is similar to that used
in \citet{Yang2004}, and is described in detail in S16 \citep[see
  also][]{Luyi2015}.

Next, we proceed to construct mock galaxy samples that have the same
survey selection effects as the SDSS DR7 (introduced in Section
\ref{sec:data}). We stack the populated simulation boxes in order to
cover the volume of SDSS DR7. We then place a virtual observer at the 
center of the stack of boxes and remove all mock galaxies that are 
located outside of the SDSS DR7 survey region under a 
$(\alpha, \delta)$-coordinate system. Each galaxy is assigned the redshift and $r$-band 
apparent magnitude according to its distance, line-of-sight velocity, 
and luminosity and selected according to the position-dependent 
magnitude limit. To mimic the position-dependent completeness, we
randomly sample each galaxy using the completeness masks provided 
by the SDSS DR7. We restrict the sample to galaxies within the redshift 
range $0.01 \leq z \leq 0.2$ and with completeness $\geq 0.7$. 
Finally, in order to have a rough estimate of the cosmic variance, 
we construct a total of 10 such mock samples by randomly rotating and 
shifting the boxes in the stack. From each mock sample, four flux-limited
subsamples are constructed using the redshift and absolute magnitude 
ranges listed in Table~\ref{subsamp}.

\subsection{Testing the Reconstruction of the Velocity Field}

We start by testing the velocity field reconstructed in the
flux-limited sample.  As mentioned in \S\ref{sec:correction}, since
the groups are distributed in redshift space, the reconstruction needs
to be iterated.  As in W12, in order to facilitate a comparison with
the real-space velocity field (in the mock data cube), we first use
two iterations with our fiducial smoothing scale of ${\rm log}(M_{\rm
  s}/\msunh) = 14.75$. Next, we apply a third iteration, this time
adopting a somewhat smaller smoothing scale of ${\rm log}(M_{\rm
  s}/\msunh) = 14.0$. This third iteration results in a weaker
suppression of the (non)linear velocities, thereby giving a larger
dynamic range over which the reconstruction of the
velocity field can be tested.

Fig.~\ref{fig:vel} shows the comparison between the true group
velocities, $v_{\rm real}$, in the simulation used to construct the
mock and the velocities $v_{\rm recon}$, obtained from the
reconstruction (using the three iterations described above).  The
slope of the best-fitting relation and the rms error
between $v_{\rm real}$ and $v_{\rm recon}$ are indicated in each
panel. Perfect reconstruction would correspond to unity slope and zero
rms. Panels in the top and middle rows show results for groups in
different halo mass bins and at different redshifts,
respectively. Reconstructed velocities are linearly correlated with
the true velocities, indicating overall success for the reconstruction
method. However, there is appreciable scatter, which increases
(weakly) with group mass and redshift.  This is mainly due to the
flux-limited nature of the sample used, which ensures that more
massive halos are located at higher redshift, where the sampling of
the density field is less accurate (mainly because $M_{\rm th}$ is
larger). In addition to the scatter, there is a systematic bias, in
that the slope of the $v_{\rm real}-v_{\rm recon}$ relation deviates
from unity. In particular, for high values of $|v_{\rm real}|$, the
corresponding $|v_{\rm recon}|$ is typically too small. This is mainly
an effect of the limited volume that is used to probe the density
field; recall that the velocity field is particularly sensitive to the
large-scale modes. In order to quantify this effect, we follow W12 and
compute for each group the `filling factor' $F_{80}$, which is defined
as the fraction of grid cell centers in a spherical volume of radius
$80 \mpch$ centered on the group.  Hence, $F_{80} \ll 1$ for a group
that is close to the edge of the survey, while groups that are located
more than $80 \mpch$ away from any survey boundary will have $F_{80}
\sim 1$.  The three panels in the bottom row of Fig.~\ref{fig:vel},
show the results for groups split by $F_{80}$, as indicated. Note that
we have arranged the split so that each of the three
subsamples contains roughly an equal number of groups. For
groups with $F_{80} \geq 0.97$, the reconstructed velocity is very
accurate; the slope is close to unity, and the scatter is relatively
small. As $F_{80}$ decreases, the slope of the correlation deviates
more strongly from unity, while the scatter increases. Hence, the main
limiting factor for the velocity reconstruction is the limiting volume
probed by the SDSS data.

\subsection{Testing the Clustering of Galaxies in Reconstructed
  Spaces}
\label{sec:test}

In order to gauge the accuracy of the correction method using the
flux-limited sample, we now compare the clustering of galaxies in the
reconstructed spaces with that in the corresponding true spaces.  The
method that we use to compute the 2PCFs is described in the Appendix.

We first start with a qualitative, visual comparison based on the 2D
2PCF $\xipi$, shown in Fig. \ref{fig:2d2pcf}. Each column
corresponds to a specific magnitude bin, as indicated at the top of
each column. From top to bottom, the different rows show the results
in different spaces, as indicated to the right of each row. In each
case, black and red contours correspond to the true and reconstructed
space, respectively.  The $\xipi$ in redshift space is clearly
anisotropic, revealing the FOG effect on small scales and the impact
of the Kaiser effect on large scales.  The panels in the middle row
demonstrate that the correction for the FOG compression (giving rise
to re-Kaiser space) is fairly successful, except for some residual FOG
effects at small projected separations.  As discussed in S16, these
shortcomings of the FOG compression arise from imperfections in the
group finder and are virtually impossible to avoid with any group
finder \citep[see][for details]{Cam2015}. In fact, for FOG
compression, there is no distinction as to whether the flux- or
volume-limited sample is used.  After correcting for both RSDs, the
$\xipi$ in re-real space (bottom row) is clearly more isotropic,
showing that the correction for the Kaiser effect is fairly accurate,
even for a flux-limited sample. Once again, the residual FOG effects
are evident, but overall,		 the method appears to correct for most of the
RSD.

A more quantitative estimate can be obtained using the one-dimensional
2PCF $\xis$. Fig. \ref{fig:xiS} compares $\xis$ in re-real space (blue
filled circles) to that in real space (solid lines). Different columns
correspond to different magnitude bins, as indicated. The upper panels
show the actual 2PCFs obtained by averaging results from all 10
mocks, while the lower panels plot $\xi_{\rm re-real}/\xi_{\rm
  real}$\footnote{Note that we plot the average of the ratios, rather
  than the ratio of the averages}. Error bars indicate the variance
among the 10 mock samples and reflect the measurement error due to
cosmic variance in an SDSS-like survey. The red dashed lines show the
results in redshift space and are shown to emphasize the magnitude of
the RSDs, as well as the success of our reconstruction method.
Clearly, the correlation functions in re-real space are in excellent
agreement with those in real space, with the vast majority of data
points being consistent with $\xi_{\rm recon}/\xi_{\rm true}=1$ within
$1 \sigma$.  For faint galaxies, the reconstructed 2PCF is
systematically underpredicted on small scales, albeit at a barely
significant level. This is a manifestation of the residual FOG effects
arising from inaccuracies in the group finder.  Overall, it is clear
that the majority of RSDs have been successfully corrected. In
particular, a comparison with Fig. 5 in S16 shows that the
reconstruction presented here based on flux-limited samples is at
least as accurate as that based on volume-limited samples.

\section{Estimation of the structure Growth Rate}
\label{sec:estimation}

We now turn to our basic goal: measuring the structure growth rate
$f\sigma_8$ using intermediate-scale clustering measurements.  It is
well known that modeling RSDs can be used to estimate the value of
$\beta \equiv f/b$ \citep[e.g.][]{Pea2001, Hawkins2003, Per2004}.
Here we present a new method that can provide simultaneous
measurements of $\xi(s)$, $f\sigma_8$ (the parameter of interest), and
$b\sigma_8$ (the bias parameter). The idea is as follows. Our
reconstruction depends (strongly) on cosmology and especially on the 
value taken by $f\sigma_8$. The reconstruction gives us both the 2PCF
$\xis$ in re-real space and the two-dimensional $\xipi$ in
re-Kaiser space (i.e., $\xipi$ with FOG compression). By comparing
$\xis$ to the (cosmology-dependent) matter-matter correlation function
on large scales, we can infer $b\sigma_8$, and thus $\beta =
f/b$. Using linear theory, we can then use $\xis$ and this value for
$\beta$ to predict the two-dimensional 2PCF $\xipi$ in the
absence of the FOG effect, which can be compared directly to $\xipi$
in re-Kaiser space obtained from our reconstruction.  Only if the
correct cosmology is used will these two correlation functions agree,
thereby giving us a handle to constrain cosmological parameters, i.e.,
the linear growth rate parameter $f\sigma_8$.

In this section, we first give a detailed description of the method,
and then we test it against mock galaxy samples.

\subsection{Methodology}
\label{sec:method2}

One of the key ingredients in our method to constrain the linear
growth rate $f\sigma_8$ is the relation between $\xipi$ and $\xis$ in
the absence of nonlinear FOG effects.
According to linear theory, developed by \cite{Kai1987} and
\cite{Ham1992}, we can define $\xipimod$ as
\begin{equation}\label{eq:ximod}
\xipimod = \xi_0(s) {\cal P}_0(\mu) +
\xi_2(s) {\cal P}_2(\mu) + \xi_4(s) {\cal P}_4(\mu)
\end{equation}
Here ${\cal P}_l(\mu)$ is the $l^{\rm th}$ Legendre polynomial, $\mu$ is the
cosine of the angle between the line of sight and the redshift-space
separation $\bm{s}$, and the angular moments can be written as
\begin{eqnarray}
\xi_0(s)&=&\left(1+\frac{2\beta}{3}+\frac{\beta^2}{5}\right)\xis \\\nonumber
\xi_2(s)&=&\left(\frac{4\beta}{3}+\frac{4\beta^2}{7}\right)\left[\xis-\bar{\xi}(s)\right]\\\nonumber
\xi_4(s)&=&\frac{8\beta^2}{35}\left[\xis+\frac{5}{2}\bar{\xi}(s)-\frac{7}{2}\hat{\xi}(s)\right]
\end{eqnarray}

with
\begin{eqnarray}
\bar{\xi}(s)&=&\frac{3}{r^3}\int_0^r{\xi_(s')s'^2ds'} \\\nonumber 
\hat{\xi}(s)&=&\frac{5}{r^5}\int_0^r{\xi_(s')s'^4ds'}.
\end{eqnarray}
Note that $\beta \equiv f/b \equiv f\sigma_8/b\sigma_8$, where $f\sigma_8$ is
the same growth rate parameter as used in the reconstruction and $b\sigma_8$ is	
the linear bias of the galaxies, which is defined by
\begin{equation}\label{eq:bias}
\xi_{\rm gg}(s) = b^2 \, \xi_{\rm mm}(s)=(b\sigma_8)^2 \, \xi_{\rm mm,\sigma_8=1.0
}(s)\,,
\end{equation}
where $\xi_{\rm gg}$ and $\xi_{\rm mm}$ are the correlation functions
of galaxies and mass on large scales, respectively.

\begin{figure}
\center
\includegraphics[width=0.5\textwidth]{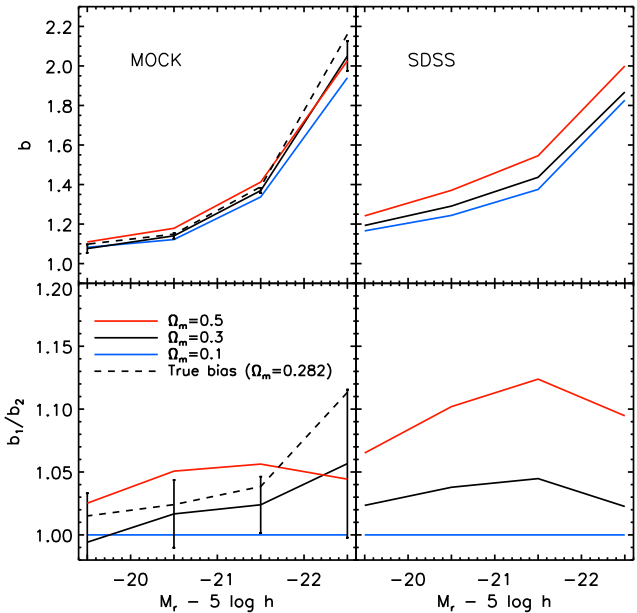}
\caption{Bias factor (top panels) and bias ratios (bottom panels)
  as a function of the absolute magnitude for mock galaxies (left
  panels) and SDSS galaxies (right panels). Here results are shown for
  a fixed $\sigma_8=0.817$. The bias factor is defined as the ratio
  of the measured re-real-space $\xis$ to that of dark matter over the
  range $4\mpch < s < 20\mpch$. Blue, black, and red solid lines
  correspond to cosmologies with $\Omega_\rmm=0.1, 0.3, 0.5$,
  respectively. The bias ratios are normalized to those for
  $\Omega_\rmm=0.1$.  The error bars correspond to the $1\sigma$ variance
  among 10 mocks. For clarity, the error bars are only plotted for the
  biases of $\Omega_\rmm=0.3$. The dashed lines in the left panels
  correspond to the true bias, as measured in our mock real space,
  which has $\Omega_\rmm = 0.282$.}
\label{fig:bias_dr7}
\end{figure} 

\begin{figure*}
\center
\includegraphics[width=0.86\textwidth]{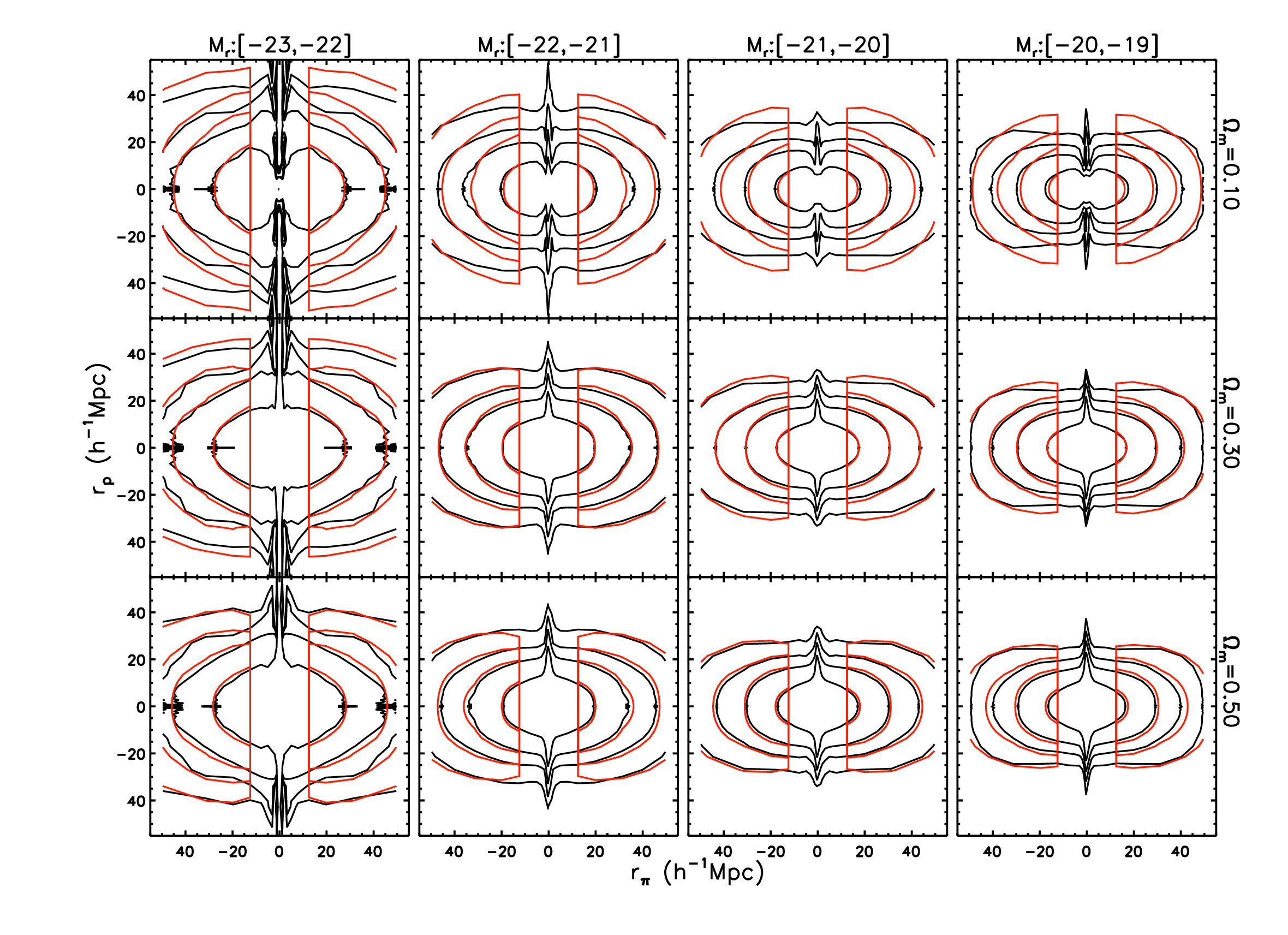}
\caption{ Validation with mocks, impact of cosmology: comparison of
  the modeled $\xipimod$ (red lines) and the measured $\xipimea$ in
  re-Kaiser space (black lines). Different rows correspond to the
  $\xipimod$ with different $\Omega_\rmm$, as indicated to the right
  of each row. Different columns correspond to different absolute
  magnitude bins, as indicated at the top of each column. Contour
  levels correspond to $\xi=0.3,~0.1,~0.05,~0.02$. Here again,  results are shown for
  a fixed $\sigma_8=0.817$.}
\label{fig:2d2pcf_3om}
\end{figure*}

In summary, our method for simultaneously constraining the growth rate
$f\sigma_8$ and the real-space 2PCF $\xi(s)$ (including bias parameter
$b\sigma_8$) consists of the following steps.
\begin{enumerate}
\item Pick a set of cosmological parameters and compute the
  corresponding value for $f\sigma_8$. In practice, we use the fitting
  function of \citet{Lah1991}:
\begin{equation}
f(z) \simeq \Omega_\rmm^{0.6}(z) + \frac{1}{70}\Omega_\Lambda(z)\left
             [1 +\frac{\Omega_\rmm(z)}{2}\right].
\end{equation}
\item Reconstruct the velocity field using the flux-limited group sample.
      \begin{enumerate}
      \item Run the group finder over the data. Since this involves
        measuring distances and absolute magnitudes, this step is
        cosmology-dependent, as are all subsequent steps below.
        \item Assign halo masses to the groups using rank-order matching onto
           the characteristic group luminosity (see \S\ref{sec:data} for details). 
        \item For each redshift range listed in Table~\ref{Subvol}, determine
           $M_{\rm th}$ and compute the corresponding $b_{\rm hm}$ using
           Eq.~(\ref{bhm}).
        \item Apply the reconstruction method, which consists of the following
           steps: (i) Fourier transform the density field $\delta(\bm{x}) =
           \delta_\rmh(\bm{x})/b_{\rm hm}$, (ii) compute $\bm{v}(\bm{k})$ using
           Eq.~(\ref{eq_vk}), and (iii) Fourier transform $\bm{v}(\bm{k})$ to
           obtain the velocity field $\bm{v}_{\rm cen}(\bm{x})$. As discussed
           in \S\ref{sec:correction}, we iterate this process until convergence
           is achieved.
        \end{enumerate}
\item Measure the 2PCF in re-Kaiser space and re-real space.
       \begin{enumerate}
         \item Apply the statistical FOG compression using the method described
            in \S\ref{sec:correction}, and compute the two-dimensional 2PCF in
            re-Kaiser space, which we denote by $\xipimea$.
         \item Correct for the Kaiser effect by reassigning galaxies their
            corrected redshifts, given by Eq.~(\ref{zcorr}). Compute the
            corresponding comoving distances and use these to compute the 2PCF
            $\xis$ in re-real space.
        \end{enumerate}
\item Estimate $\xipimod$
       \begin{enumerate}
       \item Compute the galaxy bias parameter $b\sigma_8$, using
         Eq.~(\ref{eq:bias}) applied to the separation range
         $4\mpch < s < 20\mpch$. Throughout, $\xi_{\rm mm}(s)$ is
         computed from the nonlinear matter power spectrum using the
         method of \citet{Smith2003} and adopting the transfer
         function of \citet{EH1998}.
         \item Compute $\beta = f/b$, and use this together with $\xis$ to
             compute $\xipimod$ using Eq.~(\ref{eq:ximod}). 
       \end{enumerate}
\item In order to quantify the level of agreement between $\xipimod$
  and $\xipimea$ we compute
\begin{equation}\label{Eq:Chi2}
\chi^2=\sum{\left[ \frac{\xipimod-\xipimea}{\sigma} \right]^2}.
\end{equation}
Here $\sigma$ is the rms of $\xipimea$ determined from each of
our 10 mock samples, and the summation is over a total of $4 \times
31$ logarithmic bins in $r_\rmnp$ (spanning the range $10 \mpch <
r_\rmnp < 60 \mpch$) and $r_\pi$ (spanning the range $0.06 \mpch <
r_\pi < 80 \mpch$). In order to allow for a fair comparison among
  the different sets of realizations, we always use the same
  $\sigma$, which has been obtained for the WMAP9 cosmology.

\item We repeat steps 1-5 to search for the set of cosmological
  parameters that yields the minimum $\chi^2$ value.
\end{enumerate}

Note that in computing $\chi^2$, we exclude all data with $r_\rmnp <
10\mpch$. The reason is twofold. First of all, there are still
residual FOG effects in re-Kaiser space that show up on small scales
(cf. Fig.~\ref{fig:2d2pcf}). Second, the linear theory
prediction of Eq.~(\ref{eq:ximod}) becomes inaccurate on small,
quasi-linear scales \citep[$\lta 8\mpch$;][]{Reid2014}. We find that
including data on smaller scales by reducing the lower limit of
$r_\rmnp$, increases the $\chi^2$. Focusing on larger scales by
increasing the lower limit of $r_\rmnp$ results in noisier, less
accurate constraints. Following multiple tests, we found the choice of
$r_\rmnp < 10\mpch$ to be a good compromise between these two effects.

Note that cutting results below $r_\rmnp = 10\mpch$ does not mean that
one can ignore the FOG compression when computing $\xipimea$. The FOG
effect caused by the nonlinear velocities on small scales has
significant impact on the large-scale clustering of galaxies. This is
evident from Fig. \ref{fig:2d2pcf}, which reveals clear differences
between $\xipi$ in redshift space and in Kaiser space out to large
$r_\rmnp$. Hence, it is necessary to compress the FOG effects, even when
only modeling the linear $\xipi$ on large scales
($r_\rmnp > 10 \mpch$).  We will explicitly demonstrate this in the
next section.

\begin{figure*}
\center
\includegraphics[width=0.9\textwidth]{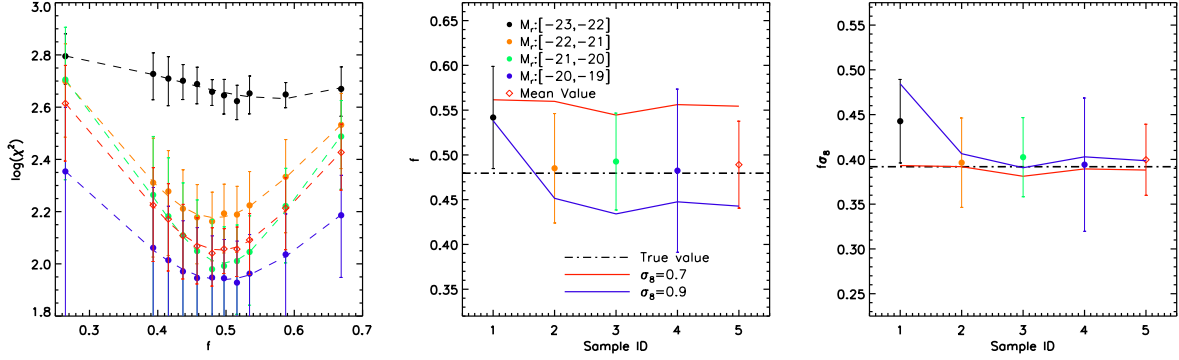}
\caption{ Validations with mocks: left panel shows $\log(\chi^2)$ as a
  function of $\fo$. This $\chi^2$ is computed using
  Eq.~(\ref{Eq:Chi2}) using the $\xipimea$ measured in re-Kaiser
  space and assuming the same value for $\sigma_8$ ($0.817$) as in
  the simulation used to construct the mocks. Filled circles show
  results for galaxies in four different magnitude bins as indicated
  with different colors, while error bars indicate the $\pm 1\sigma$
  variance among our 10 mocks.  Diamonds show the corresponding
  results averaged over the three faintest samples. Dashed lines are
  simple polynomials fit to these $\chi^2(f)$ results. The middle
  panel shows the best-fitting $f$ in different magnitude samples
  (circles) and the corresponding averaged results (diamond).  
    The sample IDs ($x$-axis, from 1 to 4) used in the middle and right
    panels correspond to the ones listed in Tabel~\ref{subsamp}, while
    sample 5 is for the average results based on the mean value of
    $\chi^2$ for the samples 2-4. The dot-dashed line indicates the
  true value of $f$ in the simulation.  The red and blue solid curves
  show the best-fitting $f$ obtained assuming $\sigma_8=0.7$ and 0.9,
  respectively. The right panel is similar to the middle panel but for
  $f\sigma_8$ values.}
\label{fig:Chi2}
\end{figure*} 
\begin{figure*}
\center
\includegraphics[width=0.9\textwidth]{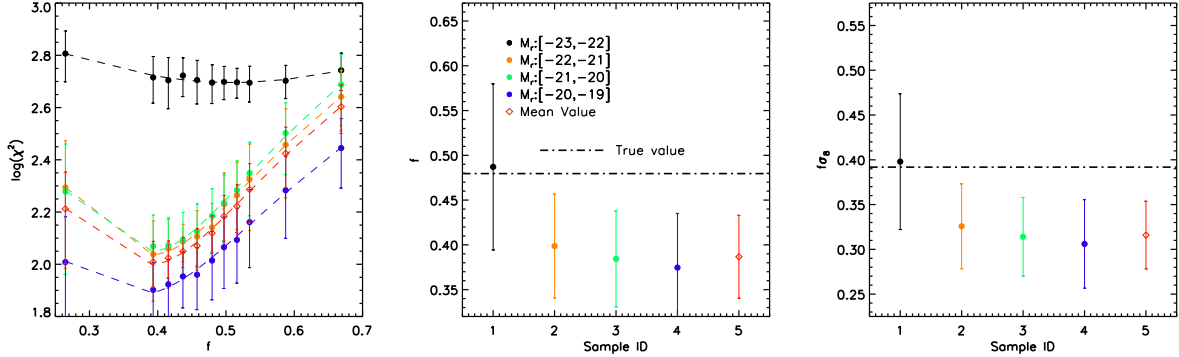}
\caption{ Validations with mocks: same as Fig. \ref{fig:Chi2} but here
  $\chi^2$ is computed using $\xipimea$ measured in redshift space,
  rather than in re-Kaiser space. Note that not including FOG
  compression results in significantly biased inferences, even when
  excluding data with $r_\rmnp < 10\mpch$}
\label{fig:Chi2_red}
\end{figure*} 

\begin{figure*}
\center
\includegraphics[width=0.86\textwidth]{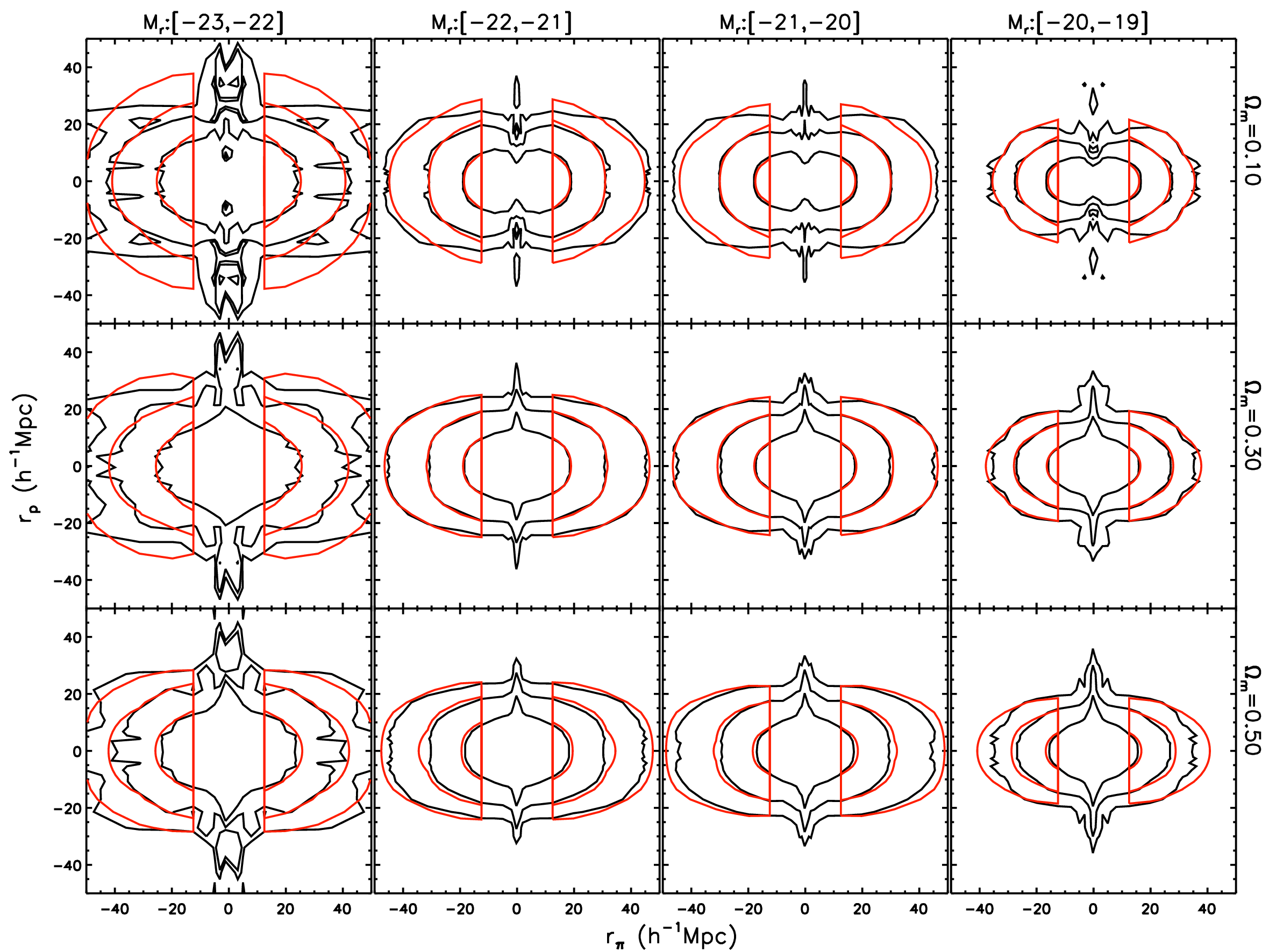}
\caption{ Application to SDSS: comparison of the modeled $\xipimod$
  (red lines) and the measured $\xipimea$ in re-Kaiser space (black
  lines) for SDSS DR7 data. Different rows correspond to the
  $\xipimod$ with different $\Omega_\rmm$, as indicated to the right
  of each row. Different columns correspond to different absolute
  magnitude bins, as indicated at the top of each column. Contour
  levels correspond to $\xi=0.3,~0.1,~0.05$.  }
\label{fig:2d2pcf_3om_dr7}
\end{figure*}

\subsection{Tests Based on Mock Data}
\label{sec:test2}

In order to test the method outlined above, we use the mock data
described in \S\ref{sec:mocks}, which is based on a cosmological
$N$-body simulation that adopts the WMAP9 cosmology with
$\Omega_\rmm= 0.282$ and $\sigma_8 = 0.817$.  The corresponding value
for $f$ is $0.48$, which we refer to as the `true' value.  Since the
velocity field depends on a combination of $f$ and $\sigma_8$, in our
investigation, we will use a fixed $\sigma_8 = 0.817$ at first and
change to other $\sigma_8$ values at the second stage.  We analyze the
mock data assuming 11 different values of $\Omega_\rmm$: $0.10$,
$0.20$, $0.22$, $0.24$, $0.26$, $0.28$, $0.30$, $0.32$, $0.34$,
$0.40$, and $0.50$. The corresponding values for $f$ range from
$0.265$ to $0.669$.  In each case, we adjust $\Omega_\Lambda$ accordingly
so as to assure a flat cosmology ($\Omega_\rmm + \Omega_\Lambda = 1$),
while all other cosmological parameters are held fixed to the WMAP9
cosmology used for the simulation.

First, we test how well the method can recover the galaxy bias
parameter $b$, and how the inferred value depends on the assumed value
of $\Omega_\rmm$. The top left panel of Fig. \ref{fig:bias_dr7}
shows the bias factor $b$ as a function of galaxy luminosity. Solid
lines indicate the results inferred from Eq.~(\ref{eq:bias}), where
$\xis$ is the correlation function of mock galaxies in the
reconstructed re-real space, and $\xi_{\rm mm}(r)$ is the nonlinear
matter 2PCF for the assumed cosmology (i.e., the assumed value of
$\Omega_\rmm$, as indicated). For comparison, the dashed line
indicates the bias factor $b$ inferred from Eq.~(\ref{eq:bias}) using
the `true' real-space 2PCF of mock galaxies and the 2PCF of the dark
matter for the actual cosmology of the simulation.  Hence, this bias
factor basically represents the true bias of the mock galaxies.  The
error bars reflect the $1\sigma$ variance among the 10 mocks and are
only plotted for the results for $\Omega_\rmm=0.3$ for clarity.  The
bottom left panel shows the ratios of the bias with respect to that
for $\Omega_\rmm =0.1$. Note that all of these results correspond to
$z=0.0$, which is the redshift of the simulation output that we used
to construct the mock data.

It is reassuring that the reconstructed real-space bias $b$ best
matches the `true' bias for $\Omega_\rmm = 0.3$, which is the value
that is closest to the actual value used in the simulation ($0.282$).
Adopting $\Omega_\rmm=0.1$ ($0.5$) results in an inferred bias that is
systematically too low (high). We thus conclude that our method can
adequately recover the bias parameter as long as the assumed cosmology
is correct, while an incorrect cosmology results in a systematic error
in the inferred bias (see also S16 for additional tests).

Next, we compare $\xipimea$ with the corresponding model prediction,
$\xipimod$. As discussed above, an incorrect cosmology results in a
value for $f$ ($f\sigma_8$) that deviates from the true value, which
in turn will introduce systematic errors in $\ximod$. Fig.
\ref{fig:2d2pcf_3om} shows the comparison between $\ximod$ (red lines)
and $\ximea$ (black lines). Different rows show the $\ximod$ inferred
for three different cosmologies, $\Omega_\rmm= 0.1,~0.3,~0.5$, from
top to bottom.  Different columns correspond to different magnitude
bins, as indicated at the top of each column. As for the bias
parameter $b$, the results for $\Omega_\rmm=0.3$, which is closest to
the real value, are in better agreement with $\ximea$, at least for
the three fainter magnitude bins. For the brightest magnitude bin, the
results for $\Omega_\rmm=0.5$ actually appear to give a better match
on large scales, but the results for this magnitude bin are quite
noisy due to the low number density of brighter galaxies.

In order to make this more quantitative, we now focus on the
$\chi^2$ values defined by Eq.~(\ref{Eq:Chi2}). The left panel of
Fig. \ref{fig:Chi2} plots the logarithm of $\chi^2$ as a function of
$f$. Filled, colored circles correspond to different absolute magnitude
bins, as indicated, while error bars reflect the $\pm 1 \sigma$
variance among the 10 mock samples. Clearly, in the three faint bins,
there is a significant, consistent minimum value for $\chi^2$,
indicating a best-fitting $f$. For the brightest magnitude bin,
though, the minimum is less pronounced and shifts to higher values of
$f$. The open diamonds show the mean value averaged over the
  three faintest bins. The middle panel of Fig. \ref{fig:Chi2}
shows the best-fitting $f$ for each of our four magnitude bins (filled
circles), obtained by fitting a polynomial to the $\chi^2(f)$ relation
(shown as dashed lines in the left panel).  The open diamond
  shows the best fit based on the mean value of $\chi^2$ for the three
  faintest bins. Error bars reflect the $1\sigma$ variance determined from
each of our 10 mocks. For comparison, the dot-dashed line indicates
the true value of $f$. Clearly, our best-fit values for $f$ are in
excellent agreement with this true value, except for the brightest
magnitude bin, which is biased toward a higher value. The mean inferred 
$f$ from the three faint bins is $0.488 \pm 0.046$. Given that the true 
value of $f$ is $0.48$, we thus conclude that our method, when applied 
to an SDSS-like survey, is able to infer an unbiased estimate of the growth
rate parameter $f$ to an accuracy of $\sim 10\%$.

Having tested the performance of our constraints on $f$ using the true
$\sigma_8$ value ($\sigma_8=0.817$), we proceed to probe the impact of
using different $\sigma_8$ values. Assuming $\sigma_8=0.7$ and $0.9$,
we perform the same procedures that we applied for our fiducial $\sigma_8
=0.817$ case to constrain the related $f$ values. The results are indicated 
by the red and blue lines in the middle panel of Fig. \ref{fig:Chi2}, which
show that lowering $\sigma_8$ systematically increases the model
prediction for $f$, and vice versa. This behavior is expected from the
fact that the velocity field is governed by $f\sigma_8$ (see
Eq.~\ref{eq_vk2}). This is confirmed by the right panel of Fig.
\ref{fig:Chi2}, which shows that our method yields predictions for the
product $f\sigma_8$ that are, to good approximation, independent of
$\sigma_8$.  Note that since we have used three different values of
$\sigma_8$, each with 10 mocks, the error bars shown here are
estimated to reflect the $1\sigma$ uncertainty of $f\sigma_8$ among
all the 30 data values.

Finally, as already alluded to in the previous subsection, it is
important to include FOG compression, even when excluding data with
$r_\rmnp < 10\mpch$. To make this evident, Fig. \ref{fig:Chi2_red}
shows the equivalent of Fig. \ref{fig:Chi2}, but this time the	$\chi^2$
is computed using $\xipimea$ measured in redshift space, rather than
re-Kaiser space. Although the brightest magnitude bin now yields
inferred $f$ and $f\sigma_8$ that are consistent with the true values,
albeit with large error bars, the best-fit values of $f$ and
$f\sigma_8$ inferred from the three fainter magnitude bins are
systematically and significantly biased toward lower value. Note 
that our sampling of $f$ is actually inadequate (it is unclear whether 
we have sampled the true minimum of $\chi^2(f)$). Consequently, if 
anything, we are likely to have overestimated the best-fit 
values for $f$ and $f\sigma_8$.

\begin{figure*}
\center
\includegraphics[width=0.9\textwidth]{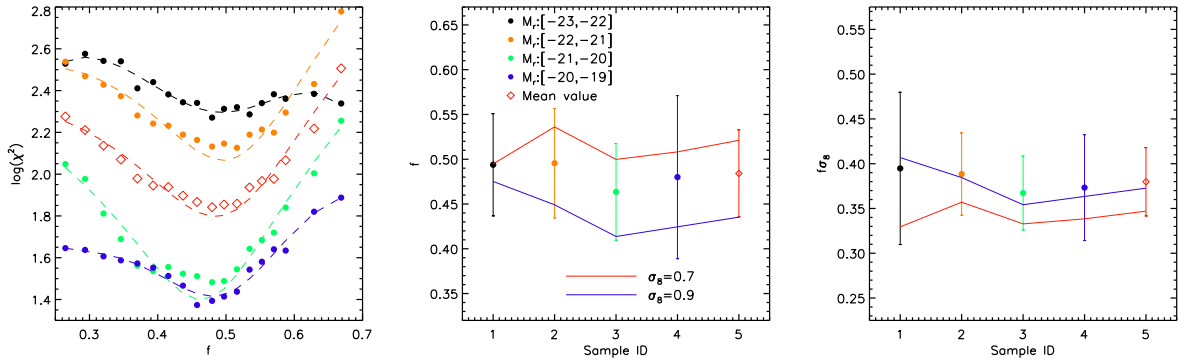}
\caption{Measurement of $f$ based on SDSS DR7 data. This figure is the
  same as Fig.~\ref{fig:Chi2}, but here we have applied our method to
  the sample of $584,473$ SDSS DR7 galaxies in the NGC (see
  Section~\ref{sec:data}).}
\label{fig:Chi2_dr7}
\end{figure*}

\section{Application to the SDSS }
\label{sec:application}

Having demonstrated that our reconstruction method is also applicable
to flux-limited samples and that we can accurately constrain the
growth rate parameter $f\sigma_8$, we now apply our method to all
galaxies in the NGC region of the SDSS DR7. As described in Section
\ref{sec:data}, this sample contains 584,473 galaxies with
$z_{\rm med}=0.1$.

\subsection{Measurement of $f\sigma_8$}
\label{sec:meas8}

We now apply our six-step iterative method, outlined in Section
\ref{sec:method2}, to the SDSS data, using a set of cosmologies (i.e.,
different $\Omega_\rmm$ values). We start by keeping the value
for $\sigma_8$ fixed to $0.817$. The right panels of Fig.
\ref{fig:bias_dr7} show the bias factor and bias ratios as a function
of the absolute magnitude for SDSS galaxies. In agreement with the
mock results, larger values for $\Omega_\rmm$ result in larger
inferred values for the bias parameter $b$.

Fig. \ref{fig:2d2pcf_3om_dr7} shows the comparison between $\ximea$
and $\ximod$ for three cosmologies with $\Omega_\rmm= 0.1$, $0.3$, and
$0.5$, in different rows. Different columns correspond to the
  four absolute magnitude bins, $\bm{[}-23.0,-22.0\bm{]}$,
$\bm{[}-22.0,-21.0\bm{]}$, $\bm{[}-21.0,-20.0\bm{]}$, and
$\bm{[}-20.0,-19.0\bm{]}$, as indicated at the top of each
column. Black and red lines correspond to $\ximea$ and $\ximod$,
respectively.  Clearly, the $\ximod$ based on $\Omega_\rmm=0.3$ is in
good agreement with the corresponding $\ximea$, while for $\Omega_\rmm
= 0.1$ and $0.5$, there are clear systematic discrepancies.

The left panel of Fig. \ref{fig:Chi2_dr7} shows $\chi^2$ as a
function of the value of $f$ for the assumed cosmology. Filled,
colored circles correspond to different absolute magnitude bins, while
the open diamonds show the mean values obtained by combining the three
faintest bins.  As for the mock data, $\chi^2(f)$ reveals clear minima
from which we can infer constraints on $f$. The middle panel of Fig.
\ref{fig:Chi2_dr7} shows the best-fitting $f$ for the different
magnitude bins. Error bars indicate the $\pm1\sigma$ variance among
the 10 mock samples described in Section \ref{sec:test2}. As one can
see, the values of the best-fitting $f$ in all four magnitude bins are
consistent with each other to better than $1\sigma$. Based on our mock
results, though, we exclude the brightest magnitude bin from our final
constraint on $f$, which we base on the mean $\chi^2$ for the three
faintest bins (indicated by the open diamond). This yields a best-fit
value for the growth rate parameter of $f(z_{\rm med}=0.1)=0.484 \pm
0.049$. As always, the error bar indicates the $1\sigma$ variance
among the 10 mock samples.

As for the mocks, we now test the impact of $\sigma_8$ on our
constraints for $f$ and $f\sigma_8$ by fixing $\sigma_8=0.7$ and 0.9,
respectively.  For each $\sigma_8$, we repeat the measurement of $f$
using the method described in Section \ref{sec:method2}. The
best-fitting $f$ values are shown in the middle panel of Fig.
\ref{fig:Chi2_dr7}. Red and blue solid curves correspond to
$\sigma_8=0.7$ and $0.9$, respectively. As for the mocks, the
values for $f$ systematically and significantly increase (decrease)
with increasing (decreasing) $\sigma_8$.  However, as shown in
Eq.~(\ref{eq_vk2}) and tested with mock samples, $f\sigma_8$ should be
related to the unique velocity field in our universe and thus be
independent of the value of $\sigma_8$ used in our analysis.  The
right panel of Fig.~\ref{fig:Chi2_dr7} shows that there is indeed
good agreement between the $f\sigma_8$ values inferred assuming
$\sigma_8=0.817$ (colored symbols) or $0.9$ (blue curve).
Assuming $\sigma_8=0.7$ (red curve) yields values for $f\sigma_8$ that
are somewhat lower, although the offset is small compared to the
measurement errors (error bars, reflecting the $\pm 1\sigma$ variance
among our 30 mock samples). Note that since the SDSS data correspond
to a median redshift $z_{\rm med}=0.1$, here we have extrapolated
$\sigma_8$ to its expected value at $z=0.1$ using
$\sigma_8(z)=D(z)\sigma_8(z=0)$, with $D(z)$ the linear growth factor
normalized to unity at $z=0$.

\begin{figure}
\center
\includegraphics[width=0.5\textwidth]{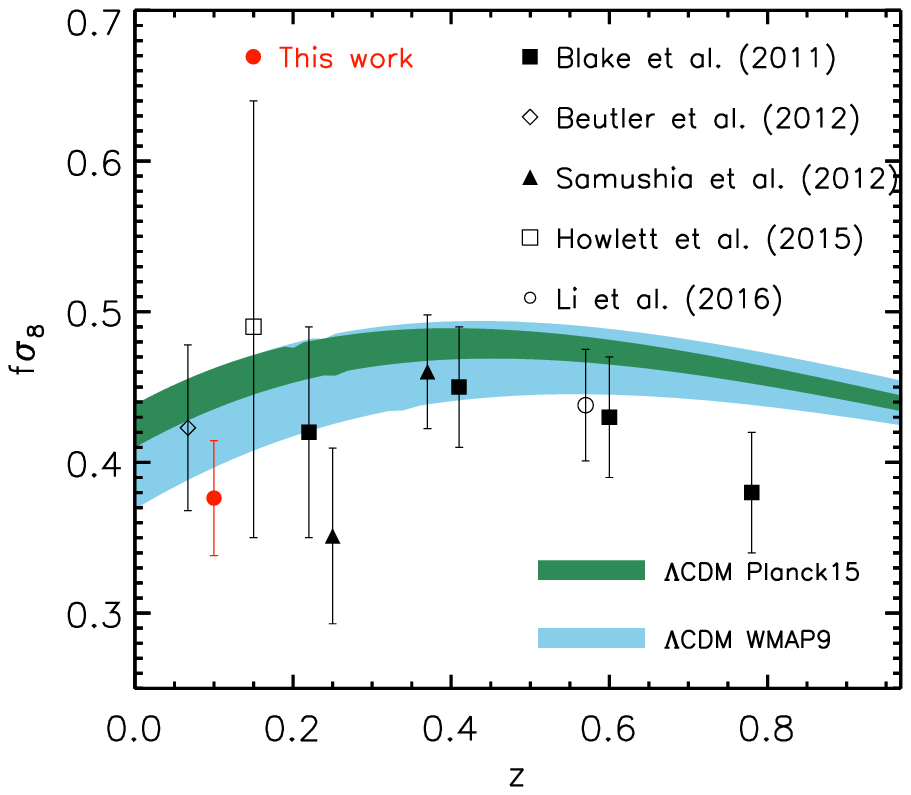}
\caption{ Comparison of the growth of structure measured at different
  redshifts. Our result is shown as the red filled circle. Black
  symbols show results from various works, including those for 6dFGS
  \citep{Beu2012}, the WiggleZ survey \citep{Blake2011}, and the SDSS
  \citep{Sam2012, How2015, Li2016}. The blue band shows the $1\sigma$
  confidence level allowed by the WMAP9 parameters assuming a flat
  $\Lambda$CDM cosmology with GR. The green band shows the $1\sigma$
  confidence level allowed by the {\it Planck} parameters
  \citep{Planck2015}, also assuming a flat $\Lambda$CDM + GR
  cosmology. }
\label{fig:fz}
\end{figure}  

Having demonstrated that our inferred value for $f\sigma_8$ is indeed
independent of the assumed value for $\sigma_8$ (at the $1\sigma$
level), we now compare our constraints on $f\sigma_8$, as inferred
under the assumption that $\sigma_8=0.817$, to constraints from
previous studies. Our results imply that $\fss$ at $z=0.1$ (open
diamond in the right panel of Fig.~\ref{fig:Chi2_dr7}).
Fig. \ref{fig:fz} compares this constraint on $f\sigma_8$ (red circle)
with previous measurements spanning a range of redshifts. These
include the results from the 6dFGS \citep{Beu2012}, the WiggleZ survey
\citep{Blake2011}, and various constraints from the SDSS
\citep{Sam2012, How2015, Li2016}. The blue band shows the $1\sigma$
confidence level allowed by the WMAP9 parameters assuming a flat
$\Lambda$CDM universe plus GR. The green band is the same but using
the {\it Planck} parameters \citep{Planck2015}.  Our measurement is
consistent with the WMAP9 prediction at the $1\sigma$ level and
somewhat lower than the {\it Planck} $\Lambda$CDM$+$GR expectations.

\begin{figure}
\center
\includegraphics[width=0.5\textwidth]{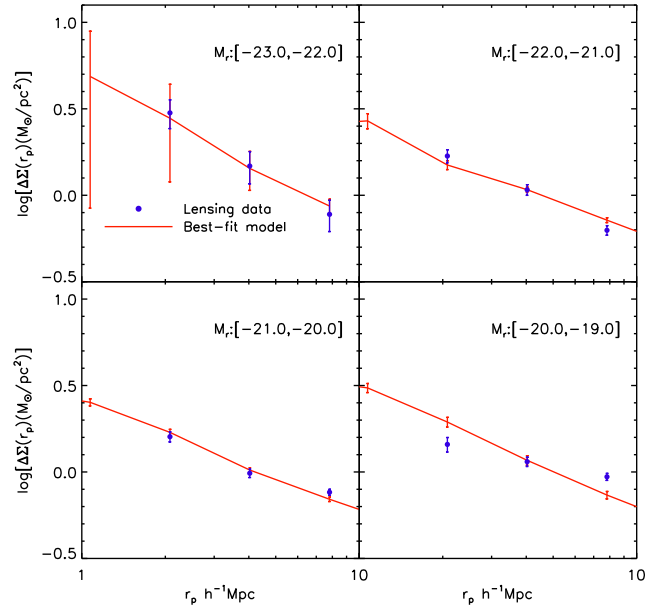}
\caption{The ESDs around lens galaxies in different absolute magnitude
  bins. The circles with error bars are the results obtained by
  \citet{Luo2017} from the galaxy-galaxy lensing shear measurements.
  The red lines are the model predictions for our best-fit
  $\Omega_{\rm m}/b$.}
\label{fig:ESD_com}
\end{figure}

\begin{figure*}
\center
\includegraphics[width=1.0\textwidth]{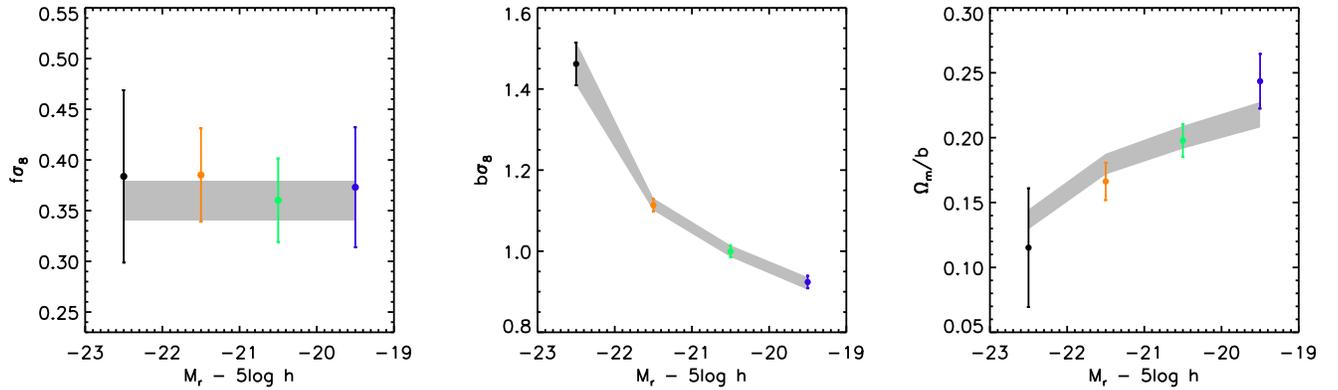}
\caption{The $f\sigma_8$ (left panel), $b\sigma_8$ (middle panel), and
  $\Omega_m/b$ (right panel) as a function of luminosity for SDSS DR7
  data. The gray band reflects the $68\%$ confidence region predicted by
  the MCMC.}
\label{fig:fs_bs}
\end{figure*}

\begin{figure*}
\center
\includegraphics[width=0.8\textwidth]{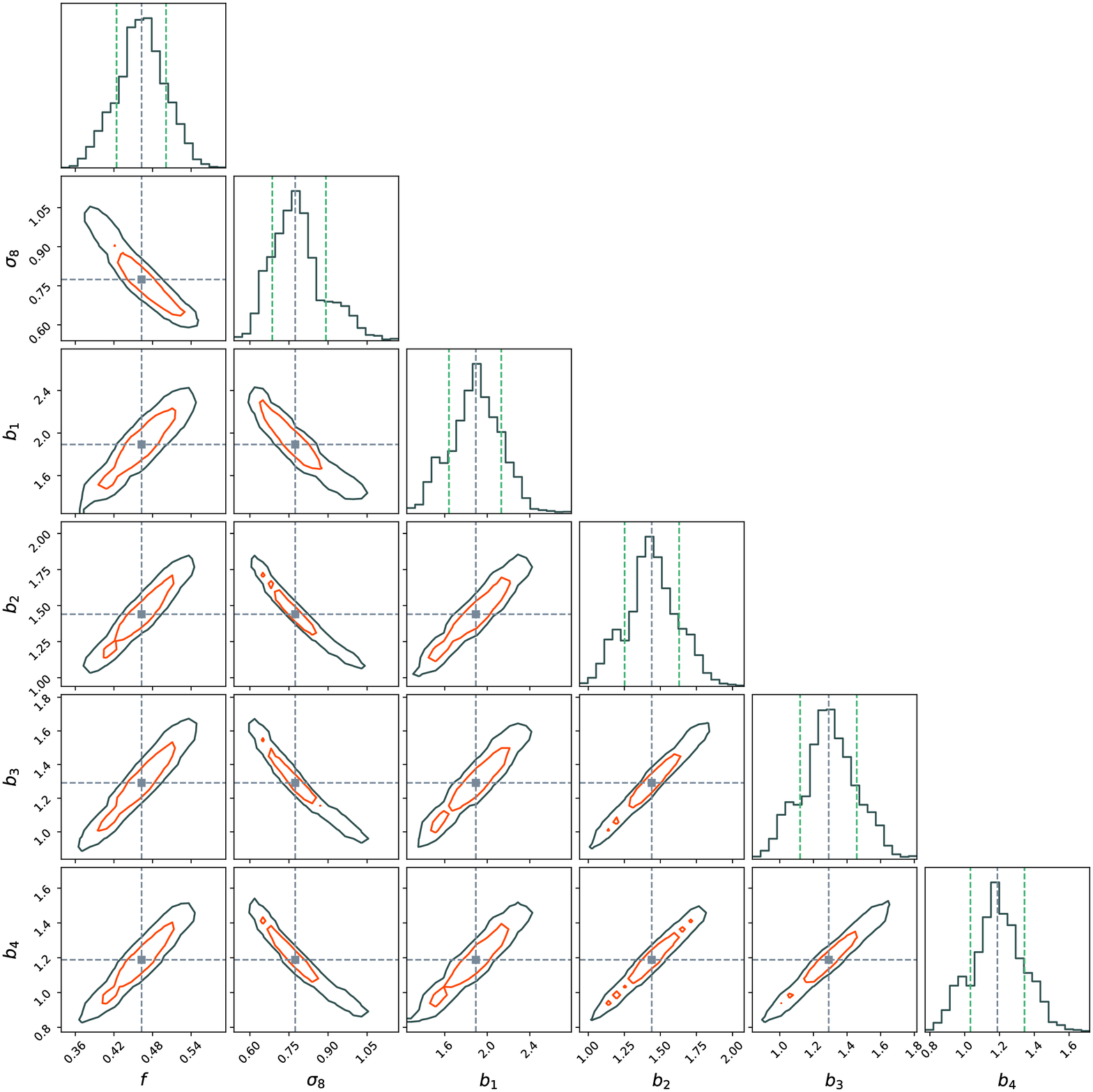}
\caption{Best fit (cross of dashed lines) and the projected
  distribution of the parameters in two-dimensional (2D) or
  one-dimensional (1D) space. The red and black contours in the 2D
  plane correspond to the boundaries of $68\%$ and $95\%$ confidence
  levels, respectively.  The 1D distributions are the marginalized
  distributions of individual parameters. The vertical black and green
  dashed lines indicate the best-fitting values and the $68\%$
  confidence region, respectively. }
\label{fig:mc2D}
\end{figure*}

\subsection{Constraints on $f$, $\sigma_8$, and $b$}

In the previous subsection, we used real-space clustering data and
RSDs to constrain $b\sigma_8$ and $f\sigma_8$, which both depend on the
value of $\sigma_8$. We now complement these data with additional
measurements that allow us to break the degeneracy among the three
parameters $f$, $\sigma_8$ and $b$. In particular, we make use of
galaxy-galaxy lensing data, which measure the excess surface density
(ESD) of galaxy lenses using shear measurements of background source
galaxies. The ESD is defined as
\begin{equation}\label{eq:ESD1}
\Delta \Sigma(r_\rmnp) = \Sigma(\leq r_\rmnp) - \Sigma(r_\rmnp) =
\gamma_{\rm t} \Sigma_{\rm crit}(z_{\rm l},z_{\rm s}) \,,
\end{equation}
where $\Sigma_{\rm crit}(z_{\rm l},z_{\rm s}) = \frac{c^2}{4\pi G}
\frac{D_{\rm s}}{D_{\rm l} D_{\rm ls}}$ is a geometry factor of the source
and lens system.  Here $\Sigma(\leq r_\rmnp)$ and $\Sigma(r_\rmnp)$
are the mean surface mass density inside of and at radius $r_\rmnp$,
respectively. The mean ESD around a lens galaxy is related to the
line-of-sight projection of the galaxy-matter cross-correlation
function, $\xi_{\rm gm}(r)$, as
\begin{equation}\label{eq:ESD2}
\Sigma(r_\rmnp) = {2\Omega_{\rm m}\rho_{\rm c}}
             \int_{r_p}^\infty \xi_{\rm gm}(r)
             \frac{r \, {\rm d}r}{\sqrt{r^2-r_p^2}} 
\end{equation}
and
\begin{equation}\label{eq:ESD3}
\Sigma(\leq r_\rmnp) = \frac{2}{r_p^2}
   \int_{0}^{r_p} \Sigma(y) ~y \, {\rm d}y 
\end{equation}
where $\rho_{\rm c}$ is the critical density of the universe. Since we
have obtained reliable measurements of $\xi_{\rm gg}(r)$ in real
space, we can predict the corresponding ESDs by rewriting
Eq.~\ref{eq:ESD2} as follows:
\begin{equation}\label{eq:ESD4}
\Sigma(r_\rmnp) = \frac{\Omega_{\rm m}}{b}{2\rho_{\rm c}}
             \int_{r_p}^\infty \xi_{\rm gg}(r)
             \frac{r \, {\rm d}r}{\sqrt{r^2-{r_p}^2}} \,,
\end{equation}
where we have made the assumption that the cross-correlation
coefficient $r = \xi_{\rm gm} / \sqrt{\xi_{\rm gg} \xi_{\rm mm}}$ is
equal to unity (on our scales of interest).
With these relations, we can use the galaxy-galaxy lensing
measurements together with the real-space 2PCF measurements to obtain
an independent measure of $\Omega_{\rm m}/b$.

In a recent study, \citet{Luo2017} used the background galaxies in the
SDSS DR7 to measure ESDs around lens galaxies that are separated into
the same luminosity bins as adopted here.  The circles with error bars
shown in Fig. \ref{fig:ESD_com} are their ESD measurements on
relatively large scales in different absolute magnitude
bins\footnote{Note as we have tested, the ESDs obtained in the lensing
  measurements are quite independent of the cosmological parameters we
  adopted.}. The error bars shown on top of the circles are estimated
using 2500 bootstrap resamplings of the lens galaxy samples, which are
quite small and in general reflect the Poisson sampling errors.

Combing these ESD measurements, $\Delta \Sigma(r_\rmnp)$, with our
measurements of the real-space 2PCF, $\xi_{\rm gg}(r)$, we now use
Eqs.(\ref{eq:ESD1}) - (\ref{eq:ESD4}) to constrain the ratio
$\Omega_{\rm m}/b$.  Since $b$ is a scale-independent linear bias
factor, which is only accurate at sufficiently large scales, we only
use the $\Delta \Sigma(r_\rmnp)$ data over the radial range $2\mpch <
r_\rmnp < 10\mpch$. On these large scales, the cross-correlation
  coefficient is also close to unity \citep[e.g.,][]{Cac2012}. We apply
this method separately to each of our four magnitude bins, the results
of which are shown in the right panel of
Fig.~\ref{fig:fs_bs}. Error bars indicate our estimated errors on
$\Omega_{\rm m}/b$, which are computed by propagating the errors on
both $\xi_{\rm gg}(r)$ and $\Delta\Sigma(r_\rmnp)$.

By combining all our constraints on $\Omega_\rmm/b$, and $f\sigma_8$, and
$b\sigma_8$ (shown, for completeness, in the left and middle panels of
Fig.~\ref{fig:fs_bs}), we now derive constraints on the related
parameters, $f$ (or $\Omega_m$) and $\sigma_8$, as well as the bias
parameters, $b_i$ for each separate luminosity bin, $i$. To do so, we
write
\begin{eqnarray}\label{eq:teq}
    f \sigma_8       = c_{i,1} \pm \sigma_{i,1} \\
  b_i \sigma_8       = c_{i,2} \pm \sigma_{i,2} \\
\frac{\Omega_m}{b_i} = c_{i,3} \pm \sigma_{i,3}
\end{eqnarray}    
where $i$ denotes the magnitude bin ($i=1,2,3,4$), and $c_{i,1}$,
$c_{i,2}$ and $c_{i,3}$ indicate the data values shown in the left,
middle, and right panels of Fig.~\ref{fig:fs_bs}, respectively,
the corresponding errors of which are $\sigma_{i,1}$, $\sigma_{i,2}$
and $\sigma_{i,3}$.
Using these 12 measurements $(c_{i,j},\sigma_{i,j})$, where $i=1,2,3,4$
and $j=1,2,3$, we constrain the six free parameters ($\sigma_8$, $f$
$b_1$, $b_2$, $b_3$ and $b_4$), using a Monte Carlo Markov chain
(MCMC) method to explore the likelihood function in the
multidimensional parameter space. The corresponding $\chi^2$ is
defined as
\begin{small}
\begin{equation}
\chi^2 \!=\!\!\!\sum_{i=1}^{4}\!\left[
  \!\left(\!\frac{f\sigma_8 \!\!- c_{i,1}}{\sigma_{i,1}}\!\right)^2 \!\!\!+ 
  \!\left(\!\frac{ b_i \sigma_8 \!\!- c_{i,2}}{\sigma_{i,2}}\!\right)^2 \!\!\!+
  \!\left(\!\frac{\Omega_m / b_i \!\!- c_{i,3}}{\sigma_{i,3}}\!\right)^2\!\right]\,\!\!\!.\!\!
\end{equation}
\end{small}
We start the MCMC from an initial guess that is consistent with the
WMAP9 cosmology and run the MCMC for 100,000 steps.  At any point in
the chain, we generate a new set of model parameters by drawing the
shifts in the six free parameters from six independent Gaussian
distributions. The Gaussian variances are tuned so that the average
acceptance rate for the new trial model is about 0.25, and we remove
the first 10,000 models in the chain to correct for the burn-in phase.
In order to suppress the correlation between neighboring models in the
chain, we thin the chain by a factor of 10. This results in a final chain
of 9000 independent models that sample the posterior distribution.
Fig.~\ref{fig:mc2D} shows the projected two-dimensional boundaries in
the parameter space. The best-fit values are indicated by the cross of
the dashed lines. The red and black contours indicate the $68\%$ and
$95\%$ confidence levels, respectively. Not surprisingly, many
parameter pairs are strongly correlated, in particular $f$ and
$\sigma_8$, as well as $b_i$ and $\sigma_8$. Fig.~\ref{fig:mc2D} also
shows the marginalized, one-dimensional distributions for each
parameter, with vertical black and green dashed lines indicating the
mean and the $68\%$ confidence regions.

As an illustration, the solid lines in Fig.~\ref{fig:ESD_com} show the
ESD model predictions for our best-fit $\Omega_{\rm m}/b$ value. The
error bars shown on top of the solid line are obtained from the
$\xi_{\rm gg}(r)$ of 30 mock data points and reflect the cosmic
variance.  The model predictions agree extremely well with the direct
measurements. In addition, the gray bands in Fig.~\ref{fig:fs_bs} show
the 68\% confidence intervals from the posterior predictions for
$f\sigma_8$ (left panel), $b\sigma_8$ (middle panel), and
$\Omega_\rmm/b$ (right panel). Overplotted in color are our
observational constraints for each of the four magnitude bins.  As is
evident, the posterior predictions are in good agreement with these
constraints, indicating that the constraints are mutually consistent
with each other and with a $\Lambda$CDM + GR cosmology.  Combining
RSDs with weak lensing data, we have thus been able to put successful
constraints on the logarithmic derivative of the linear growth rate,
$f$, on the clustering amplitude of matter, $\sigma_8$, and on the galaxy bias
factor, $b$, for galaxies in four luminosity bins at a median redshift
$z_{\rm med}=0.1$. For reference, Table~\ref{con_para} lists the
best-fit parameters together with their $68\%$ confidence levels.

\begin{table}
\center
\scalebox{0.87}{
\begin{threeparttable}[c]
\caption{The Best-fit Parameter}\label{con_para}
\setlength{\tabcolsep}{2pt}
\begin{tabular}{cccccc}
\toprule

\multirow{2}{*}{$f$ }  &
\multirow{2}{*}{$\sigma_8$}  &
\multirow{2}{*}{$b_1$}  &
\multirow{2}{*}{$b_2$}  &
\multirow{2}{*}{$b_3$}  &
\multirow{2}{*}{$b_4$}  \\\\

\multirow{2}{*}{\f} &
\multirow{2}{*}{\s} &
\multirow{2}{*}{\bI} &
\multirow{2}{*}{\bII} &
\multirow{2}{*}{\bIII} &
\multirow{2}{*}{\bIV} \\\\

\\ \bottomrule
\end{tabular}
\textbf{Notes.}  All of the best-fit parameters listed in this table
  correspond to the values at redshift $z_{\rm med}=0.1$.  Note that
  $\sigma_8$ can be extrapolated to the value at $z=0$ using
  $\sigma_8(0)=\sigma_8(z=0.1)/D(z=0.1)$, with $D(z)$ the linear
  growth factor normalized to unity at $z=0$. The linear bias
  parameters $b_1$, $b_2$, $b_3$ and $b_4$ correspond to galaxies in
  absolute magnitude bins
  $\rmag=[-23,0, -22.0], [-22,0, -21.0], [-21.0, -20.0]$, and
  $[-20.0, -19.0]$, respectively.
\end{threeparttable}}
\end{table}

\section{Summary}
\label{sec:con}

In S16, we presented a new, reliable method to correct the RSDs in
galaxy redshift surveys and successfully applied it to the SDSS DR7
data to construct a real-space version of the main galaxy catalog.
This allows for an accurate, `direct' measurement of the real-space
correlation function. In this paper, the second in a series, we use
the reconstructed galaxy distribution to constrain $f$ and $\sigma_8$,
as well as the linear galaxy bias parameter, $b$, in different
luminosity bins. Here $f = \rmd\ln D/\rmd\ln a$ is the logarithmic
derivative of the linear growth factor, $D$, with respect to the scale
factor, $a$, and $\sigma_8$ is the clustering amplitude of matter.

We first extended our reconstruction method so that it can be applied
to flux-limited, rather than volume-limited, samples of galaxies.
This significantly increases both the number of galaxies available
and the volume being probed, thereby improving the overall accuracy.
Using a suite of 10 mock SDSS DR7 galaxy catalogs, we tested the
performance of our RSD-correction method by comparing the two-point
clustering statics in different spaces. We have shown that the
clustering in our reconstructed re-real space is in good agreement
with that in the corresponding real space. This indicates that our
method works well, and thus that we can accurately correct for RSDs in
flux-limited samples, allowing for an accurate, unbiased measurement
of the real-space correction function.

Using this reconstruction technique, we have developed a method to
constrain the growth of the structure parameter $f$, the amplitude of
fluctuation $\sigma_8$, and the galaxy bias parameter $b$, using
clustering measurements of galaxies on intermediate scales. Our method
works as follows.
\begin{enumerate}
\item Using the 2PCF $\xis$ in reconstructed re-real space, which is
  cosmology-dependent, infer the galaxy bias parameter $b\sigma_8$ by
  comparing $\xis$ to the (cosmology-dependent) matter-matter
  correlation function.
\item Using the value for $f\sigma_8$ used in the reconstruction,
  evaluate the parameter $\beta = f/b$. Use this, in combination with
  $\xis$, to predict $\xipimod$ based on linear theory.
\item Compare $\xipimod$ to the 2D 2PCF $\xipimea$ inferred directly
  from the redshift-space distribution of galaxies after applying a
  FOG compression based on a galaxy group catalog. These two
  measurements will only agree if the correct cosmology is
  adopted. Note that failing to apply this FOG compression results in
  significant systematic errors in the inferred cosmological
  parameters (e.g., $f\sigma_8$), even when excluding all data on
  scales $r_\rmnp < 10\mpch$.
\item Use the measurements of the 2PCF in re-real space, $\xis$,
  together with measurements of the ESD,
  $\Delta\Sigma(r_\rmnp)$ of the same galaxies as inferred from
  galaxy-galaxy weak lensing, to constrain the ratio $\Omega_\rmm/b$.
\item Combine the constraints on $f\sigma_8$, $b\sigma_8$ and
  $\Omega_m/b$, to constrain $f$, $\sigma_8$ and the bias parameter,
  $b$, for each separate luminosity bin.
\end{enumerate}

Using realistic mock samples, we have shown that this method, when
applied to an SDSS-like survey, can yield an unbiased estimate of
$f\sigma_8$, with a statistical error of $\sim 10\%$.  When applying
this method to the SDSS DR7, we obtained $\fss$ at $z=0.1$.  This
value is consistent (within the $1\sigma$ level) with the $\Lambda$CDM
cosmology with WMAP9 parameters, but in slight tension (at the $\sim 2\sigma$
level) with the parameters advocated by the {\it Planck} mission.
            
By combining the clustering of galaxies measured in the re-real and
re-Kaiser spaces with galaxy-galaxy weak lensing measurements for the
same sets of galaxies, we obtain the following set of cosmological
constraints at a median redshift $z=0.1$: $f=$ \f , and $\sigma_8=$ \s
.  In addition, we are able to constrain the linear bias parameter of
galaxies in absolute magnitude bins $\rmag=[-23,0, -22.0], [-22,0,
  -21.0], [-21.0, -20.0]$, and $[-20.0, -19.0]$ to $b=$ \bI, \bII,
\bIII, and \bIV, respectively.


\section*{Acknowledgments}

We thank the anonymous referee for helpful comments that improved the
presentation of this paper.  This work is supported by the 973 Program
(No. 2015CB857002), National Science Foundation of China (grant
Nos. 11233005, 11421303, 11522324, 11503064, 11621303, 11733004) and
Shanghai Natural Science Foundation, Grant No.  15ZR1446700. We are also
thankful for the support of the Key Laboratory for Particle Physics,
Astrophysics and Cosmology, Ministry of Education.  HJM would like to
acknowledge the support of NSFC-11673065 and NSF AST-1517528, and FvdB
is supported by the US National Science Foundation through grant AST
1516962.

A computing facility award on the PI cluster at Shanghai Jiao Tong
University is acknowledged. This work is also supported by the High
Performance Computing Resource in the Core Facility for Advanced
Research Computing at Shanghai Astronomical Observatory.


\appendix

\section{The 2PCF}
\label{sec:2pf}

The two-dimensional 2PCF, $\xi(r_{\rm p},r_\pi)$, is computed
using the following estimator\citep{Ham1993}.
\begin{equation}\label{eq:2pcf}
\xi(r_\rmnp,r_\pi) = \frac{\langle RR \rangle
\langle DD \rangle}{\langle DR \rangle^2} - 1\,,
\end{equation}
where $\langle DD \rangle$, $\langle RR \rangle$ and $\langle DR
\rangle$ are, respectively, the number of galaxy-galaxy, random-random
and galaxy-random pairs with separation $(r_\rmnp,r_\pi)$.  
The variables $r_\rmnp$ and $r_\pi$ are the pair
separations perpendicular and parallel to the line of sight,
respectively. Explicitly, for a pair of galaxies, one located at $s_1$
and the other at $s_2$,   where $s_i$ is computed using
\begin{equation}
\label{distance_z}
s(z)=\frac{1}{H_0}\int_0^z
\frac{{\rm d}z}{\sqrt[]{\Omega_\Lambda+\Omega_\rmm(1+z)^3}},
\end{equation}
then we define
\begin{equation}
r_\pi =\frac{\bm{s}\cdot\bm{l}}{\bm{\mid l \mid}},
\quad r_\rmnp = \sqrt{\bm{s} \cdot \bm{s}-r_{\pi}^2}\,.
\end{equation}
Here $\bm{l} = (\bm{s_1}+\bm{s_2})/2$ is the line of sight
intersecting the pair and $\bm{s}=\bm{s_1}-\bm{s_2}$.

The one-dimensional, redshift-space 2PCF, $\xi(s)$, is estimated by
averaging $\xi(r_\rmnp,r_{\pi})$ along constant $s=\sqrt{r^2_{\rm
    p}+r^2_{\pi}}$ using
\begin{equation}\label{eq:xir}
  \xi(s) = \frac{1}{2} \, \int_{-1}^{1}{\xi(r_\rmnp,r_{\pi}) \,
    {\rm d}\mu}\,,
\end{equation}
where $\mu$ is the cosine of the angle between the line of sight and
the redshift-space separation vector $\bm{s}$.  Alternatively, one can
also measure $\xi(s)$ by directly counting $\langle DD \rangle$,
$\langle RR \rangle$, and $\langle DR \rangle$ pairs as a function of
redshift-space separation $s$.




\begin{thebibliography}{}

\bibitem[Abazajian et al.(2009)]{2009ApJS..182..543A} Abazajian, K.~N., 
Adelman-McCarthy, J.~K., Ag{\"u}eros, M.~A., et al.\ 2009, \apjs, 182, 543 


\bibitem[Alam et al.(2015)]{Alam2015} Alam, S., Ho, S., Vargas-Maga{\~n}a, 
  M., \& Schneider, D.~P.\ 2015, \mnras, 453, 1754 


\bibitem[Amendola et al.(2005)]{Amen2005} Amendola, L., Quercellini,
  C., \& Giallongo, E.\ 2005, \mnras, 357, 429 


\bibitem[Beutler et al.(2012)]{Beu2012} Beutler, F., Blake, C., 
  Colless, M., et al.\ 2012, \mnras, 423, 3430 

\bibitem[Beutler et al.(2014)]{Beu2014} Beutler, F., 
  Saito, S., Seo, H.-J., et al.\ 2014, \mnras, 443, 1065 


\bibitem[Blake et al.(2011)]{Blake2011} Blake, C., Brough, S., 
  Colless, M., et al.\ 2011, \mnras, 415, 2876 

\bibitem[Blanton et al.(2005)]{2005AJ....129.2562B} Blanton, M.~R., Schlegel, 
D.~J., Strauss, M.~A., et al.\ 2005, \aj, 129, 2562 


\bibitem[Cacciato et al.(2012)]{Cac2012} Cacciato, M., Lahav O., van den 
  Bosch, F.~C., Hoekstra H., Dekel A., 2012, \mnras, 426, 566
  
\bibitem[Cacciato et al.(2013)]{Cac2013} Cacciato, M., van den 
Bosch, F.~C., More, S., Mo, H.~J., \& Yang, X.\ 2013, \mnras, 430, 767

\bibitem[Cai \& Bernstein(2012)]{Cai2012} Cai, Y.-C., \& Bernstein, 
 G.\ 2012, \mnras, 422, 1045 


\bibitem[Campbell et al.(2015)]{Cam2015} Campbell, D., van den Bosch, F.~C.,
  Hearin, A., Padmanabhan, N., Berlind, A., Mo, H.~J.; Tinker, J., \& Yang, X.\
  2015, \mnras, 452, 444

\bibitem[Chuang et al.(2013)]{Chuang2013} Chuang, C.-H., Prada, F., 
  Cuesta, A.~J., et al.\ 2013, \mnras, 433, 3559 


\bibitem[Davis \& Peebles(1983)]{Dav1983} Davis, M., \& Peebles, P.~J.~E.\ 
1983, \apj, 267, 465 


\bibitem[Davis et al.(1985)]{Dav1985} Davis, M., Efstathiou, 
G., Frenk, C.~S., \& White, S.~D.~M.\ 1985, \apj, 292, 371 

\bibitem[de la Torre et al.(2013)]{Torre2013} de la Torre, S., 
 Guzzo, L., Peacock, J.~A., et al.\ 2013, \aap, 557, A54 


\bibitem[Eisenstein \& Hu(1998)]{EH1998} Eisenstein, D.~J., 
 \& Hu, W.\ 1998, \apj, 496, 605 

\bibitem[Hamilton(1992)]{Ham1992} Hamilton, A.~J.~S.\ 1992, \apjl,
  385, L5

\bibitem[Hamilton(1993)]{Ham1993} Hamilton, A.~J.~S.\ 1993, \apj, 417,
  19


\bibitem[Hawkins et al.(2003)]{Hawkins2003} Hawkins, E.,
  Maddox, S., Cole, S., et al.\ 2003, \mnras, 346, 78 


\bibitem[Hinshaw et al.(2013)]{Hin2013} Hinshaw, G., Larson, 
D., Komatsu, E., et al.\ 2013, \apjs, 208, 19


\bibitem[Howlett et al.(2015)]{How2015} Howlett, C., Ross, 
   A.~J., Samushia, L., Percival, W.~J., \& Manera, M.\ 2015, \mnras, 449, 848 

\bibitem[Jain \& Zhang(2008)]{Jain2008} Jain, B., \& Zhang, P.\ 2008, 
  \prd, 78, 063503 

\bibitem[Jackson(1972)]{Jac1972} Jackson, J.~C.\ 1972, \mnras, 156, 1P 

\bibitem[Jennings et al.(2011)]{Jen2011} Jennings, E., Baugh, C.~M., 
 \& Pascoli, S.\ 2011, \apjl, 727, L9 

\bibitem[Kaiser(1987)]{Kai1987} Kaiser, N.\ 1987, \mnras, 227, 1 


\bibitem[Lahav et al.(1991)]{Lah1991} Lahav O., Lilje P. B., Primack J. R., 
Rees M. J. \ 1991, \mnras, 251, 128


\bibitem[Li et al.(2016)]{Li2016} Li, Z., Jing, Y.~P., Zhang, P., \& 
  Cheng, D.\ 2016, \apj, 833, 287 

\bibitem[Linder \& Cahn(2007)]{LinderCahn2007} Linder, E.~V., \&
  Cahn, R.~N.\ 2007, Astroparticle Physics, 28, 481 

\bibitem[Linder(2008)]{Linder2008} Linder, E.~V.\ 2008, 
  Astroparticle Physics, 29, 336 


\bibitem[Lu et al.(2015)]{Luyi2015} Lu, Y., Yang, X., \& Shen, S.\ 2015, 
  \apj, 804, 55 

\bibitem[Luo et al.(2017)]{Luo2017} Luo, W., Yang, X., Zhang, 
 J., et al.\ 2017, \apj, 836, 38 




\bibitem[Navarro et al.(1997)]{NFW1997} Navarro, J. F., Frenk, C. S., \&
  White, S. D. M.\ 1997, \apj, 490,493 

\bibitem[Oka et al.(2014)]{Oka2014} Oka, A., Saito, S., Nishimichi, 
  T., Taruya, A., \& Yamamoto, K.\ 2014, \mnras, 439, 2515 


\bibitem[Peacock et al.(2001)]{Pea2001} Peacock, J.~A., Cole, 
  S., Norberg, P., et al.\ 2001, \nat, 410, 169 

\bibitem[Percival et al.(2004)]{Per2004} Percival, W.~J., 
  Burkey, D., Heavens, A., et al.\ 2004, \mnras, 353, 1201 


\bibitem[Percival \& White(2009)]{Per2009} Percival, W.~J., 
 \& White, M.\ 2009, \mnras, 393, 297 

\bibitem[Planck Collaboration et al.(2015)]{Planck2015} Planck Collaboration, 
Ade, P.~A.~R., Aghanim, N., et al.\ 2015, arXiv:1502.01589 


\bibitem[Regos \& Geller(1991)]{Reg1991} Regos, E., \& Geller, M.~J.\ 1991, 
  \apj, 377, 14

\bibitem[Reid et al.(2014)]{Reid2014} Reid, B.~A., Seo, H.-J., 
  Leauthaud, A., Tinker, J.~L., \& White, M.\ 2014, \mnras, 444, 476 


\bibitem[Samushia et al.(2012)]{Sam2012} Samushia, L., 
Percival, W.~J., \& Raccanelli, A.\ 2012, \mnras, 420, 2102

\bibitem[Samushia et al.(2014)]{Sam2014} Samushia, L., 
  Reid, B.~A., White, M., et al.\ 2014, \mnras, 439, 3504 


\bibitem[Sargent \& Turner(1977)]{Sar1977} Sargent, 
W.~L.~W., \& Turner, E.~L.\ 1977, \apjl, 212, L3 


\bibitem[Shi et al.(2016)]{Shi16} Shi, F., Yang, X., 
  Wang, H., et al.\ 2016, \apj, 833, 241

\bibitem[Smith et al.(2003)]{Smith2003} Smith, R.~E., Peacock, 
J.~A., Jenkins, A., et al.\ 2003, \mnras, 341, 1311 


\bibitem[Song \& Percival(2009)]{Song2009} Song, Y.-S., 
  \& Percival, W.~J.\ 2009, JCAP, 10, 004 



\bibitem[Springel(2005)]{Spr2005} Springel, V.\ 2005, \mnras, 364, 1105


\bibitem[Tinker et al.(2008)]{Tinker2008} Tinker, J., Kravtsov, A.~V.,
  Klypin, A., et al.\ 2008, \apj, 688, 709-728

\bibitem[Tully \& Fisher(1978)]{Tul1978} Tully, R.~B., 
\& Fisher, J.~R.\ 1978, Large Scale Structures in the Universe, 79, 31 


\bibitem[Wang(2008)]{Wang2008} Wang, Y.\ 2008, JCAP, 05, 021 



\bibitem[van de Weygaert \& van Kampen(1993)]{Wey1993} van de Weygaert, R., 
 \& van Kampen, E.\ 1993, \mnras, 263, 481


\bibitem[Wang et al.(2009)]{WH2009} Wang, H., Mo, H.~J., Jing, 
   Y.~P., et al.\ 2009, \mnras, 394, 398 

\bibitem[Wang et al.(2012)]{WH2012} Wang, H., Mo, H.~J., Yang, X., \& van
  den Bosch, F.~C.\ 2012, \mnras, 420, 1809

\bibitem[White et al.(2009)]{White2009} White, M., Song, Y.-S., 
  \& Percival, W.~J.\ 2009, \mnras, 397, 1348 


\bibitem[Yang et al.(2003)]{Yang2003} Yang, X., Mo, H.~J., 
\& van den Bosch, F.~C.\ 2003, \mnras, 339, 1057 

\bibitem[Yang et al.(2004)]{Yang2004} Yang, X., Mo, H.~J., Jing, 
Y.~P., van den Bosch, F.~C., \& Chu, Y.\ 2004, \mnras, 350, 1153

\bibitem[Yang et al.(2005)]{Yang2005} Yang, X., Mo, H.~J., van den Bosch,
  F.~C., \& Jing, Y.~P.\ 2005, \mnras, 356, 1293

\bibitem[Yang et al.(2007)]{Yang2007} Yang, X., Mo, H.~J., van den Bosch,
  F.~C., et al.\ 2007, \apj, 671, 153

\bibitem[Yang et al.(2012)]{Yang2012} Yang, X., Mo, H.~J., van 
den Bosch, F.~C., Zhang, Y., \& Han, J.\ 2012, \apj, 752, 41 







  

























  























 













\end{thebibliography}
\end{document}